\documentclass[nofootinbib,aps,prl,reprint,groupedaddress,superscriptaddress,showpacs]{revtex4-2}
\usepackage[english]{babel}
\usepackage[utf8x]{inputenc}
\usepackage{amssymb}
\usepackage{amsthm}
\usepackage{amsmath}
\usepackage{hyperref}
\usepackage{subfigure}
\usepackage[percent]{overpic} %scrivo sopra le immagini
\usepackage{picture}
\usepackage{color,soul}
\usepackage{nth}
\usepackage{gensymb}
\usepackage{cleveref}
\usepackage{textcomp}
\usepackage{makecell}
\usepackage{lineno} %serve package texlive-humanities!
\usepackage{textcomp}

\begin{document}

% \preprint{APS/PRL}

% \title{Suppression of Dielectric Wakefields in Plasma Wakefield Accelerators}
\title{Experimental Observation of Space-Charge Field Screening of a Relativistic Particle Bunch in Plasma}

\author{L.~Verra}
\email{{livio.verra@lnf.infn.it}}
\affiliation{INFN Laboratori Nazionali di Frascati, Via Enrico Fermi 54, 00044 Frascati, Italy}
\author{M.~Galletti}
\affiliation{University of Rome Tor Vergata, Via Della Ricerca Scientifica 1, 00133 Rome, Italy}
\affiliation{INFN Tor Vergata, Via Della Ricerca Scientifica 1, 00133 Rome, Italy}
\affiliation{NAST Center, Via Della Ricerca Scientifica 1, 00133 Rome, Italy}
\author{R.~Pompili}
\affiliation{INFN Laboratori Nazionali di Frascati, Via Enrico Fermi 54, 00044 Frascati, Italy}
\author{A.~Biagioni}
\affiliation{INFN Laboratori Nazionali di Frascati, Via Enrico Fermi 54, 00044 Frascati, Italy}
\author{M.~Carillo}
\affiliation{University of Rome Sapienza, Piazzale Aldo Moro 5, 00185 Rome, Italy}
\author{A.~Cianchi}
\affiliation{University of Rome Tor Vergata, Via Della Ricerca Scientifica 1, 00133 Rome, Italy}
\affiliation{INFN Tor Vergata, Via Della Ricerca Scientifica 1, 00133 Rome, Italy}
\affiliation{NAST Center, Via Della Ricerca Scientifica 1, 00133 Rome, Italy}
\author{L.~Crincoli}
\affiliation{INFN Laboratori Nazionali di Frascati, Via Enrico Fermi 54, 00044 Frascati, Italy}
\author{A.~Curcio}
\affiliation{INFN Laboratori Nazionali di Frascati, Via Enrico Fermi 54, 00044 Frascati, Italy}
% \author{A.~Del Dotto}
% \affiliation{INFN Laboratori Nazionali di Frascati, Via Enrico Fermi 54, 00044 Frascati, Italy}
\author{F.~Demurtas}
\affiliation{University of Rome Tor Vergata, Via Della Ricerca Scientifica 1, 00133 Rome, Italy}
\author{G.~Di Pirro}
\affiliation{INFN Laboratori Nazionali di Frascati, Via Enrico Fermi 54, 00044 Frascati, Italy}
\author{V.~Lollo}
\affiliation{INFN Laboratori Nazionali di Frascati, Via Enrico Fermi 54, 00044 Frascati, Italy}
\author{G.~Parise}
\affiliation{University of Rome Tor Vergata, Via Della Ricerca Scientifica 1, 00133 Rome, Italy}
\author{D.~Pellegrini}
\affiliation{INFN Laboratori Nazionali di Frascati, Via Enrico Fermi 54, 00044 Frascati, Italy}
\author{S.~Romeo}
\affiliation{INFN Laboratori Nazionali di Frascati, Via Enrico Fermi 54, 00044 Frascati, Italy}
\author{G.J.~Silvi}
\affiliation{University of Rome Sapienza, Piazzale Aldo Moro 5, 00185 Rome, Italy}
\author{F.~Villa}
\affiliation{INFN Laboratori Nazionali di Frascati, Via Enrico Fermi 54, 00044 Frascati, Italy}
% \author{A.~Zigler}
% \affiliation{Racah Institute of Physics, Hebrew University, 91904 Jerusalem, Israel}
\author{M.~Ferrario}
\affiliation{INFN Laboratori Nazionali di Frascati, Via Enrico Fermi 54, 00044 Frascati, Italy}

%\date{\today}

\begin{abstract}

The space-charge field of a relativistic charged bunch propagating in plasma is screened due to the presence of mobile charge carriers. 
We experimentally investigate such screening by measuring the effect of dielectric wakefields driven by the bunch in a uncoated dielectric capillary where the plasma is confined.
We show that the plasma screens the space-charge field and therefore suppresses the dielectric wakefields when the distance between the bunch and the dielectric surface is much larger than the plasma skin depth. 
Before full screening is reached, the effects of dielectric and plasma wakefields are present simultaneously.

\end{abstract}

\maketitle
\par \textit{Introduction.---} 
% \section{Introduction}
% The ability to screen electromagnetic fields is one of the key characteristics of plasma~\cite{CHEN:BOOK}.
One of the fundamental properties of plasma is the ability to screen electromagnetic fields, due to the presence of mobile charge carriers (plasma electrons and ions)~\cite{CHEN:BOOK}.
The propagation of electromagnetic waves in plasma has been extensively studied~\cite{OSTUKA:2015}, especially in the context of plasma generation~\cite{SHINOHARA:2002}, heating and confinement~\cite{SHOJI:1980}, requiring complex and dedicated experimental setups~\cite{WHELAN:1981}.
In the case of a relativistic charged particle bunch travelling through a plasma with density $n_{pe}$, the plasma electrons rapidly move to compensate for the almost purely transverse space-charge field of the bunch. 
This effect is the cause of phenomena such as plasma wakefields~\cite{CHEN:1985,KENIGS:1987}. 

\par Because of the screening operated by the plasma, the space-charge field of the bunch is exponentially damped with the radial distance $r$ as $E_r\propto e^{-r/\delta}$~\cite{JACKSON:BOOK}, where the characteristics length $\delta = c \sqrt{m_e \varepsilon_0 / n_{pe} e^2}$ is the plasma skin depth ($c$ the speed of light, $\varepsilon_0$ the vacuum permittivity, $m_e$ and $e$ the electron rest mass and charge, respectively). 
Hence, the plasma effectively screens the space-charge field of the bunch when $r\gg \delta$. 
Despite the physics of the process being well known, no direct experimental evidence of plasma screening has been provided so far for the fields generated by a relativistic charged particle bunch.

\par In this Letter, we present a non-invasive measurement of plasma screening of the space-charge field of a relativistic electron bunch. 
We show that screening causes the suppression of dielectric wakefields in an uncoated dielectric capillary, where the plasma is generated, when the distance between the bunch and the capillary surface is much larger than the plasma skin depth.
Moreover we observe, for the first time to our knowledge, the simultaneous effect of dielectric and plasma wakefields, when full plasma screening is not achieved.

\par A relativistic, charged bunch drives wakefields when its space-charge field interacts with a slow-wave structure or medium with a large index of refraction.
%: the space-charge electric field lines end on the material surface charges and, due to the finite resistivity of the material, these charges lag slightly behind it.
Consequently, longitudinal and transverse dielectric wakefields (also known as Cherenkov wakefields) are generated~\cite{BANE:1984,GAI:1988,KEINIGS:1989,NG:1990}. 
Dielectric transverse wakefields are axis-symmetric with respect to the center of the dielectric structure. Therefore, they induce a dipolar effect when the bunch trajectory is not aligned to the longitudinal axis of the structure~\cite{PARK:2000,BATURIN:2014,BETTONI:2016,CRAIEVICH:2017,OSHEA:2020,SAVELIEV:2022}. 
The transverse wake potential is calculated as the convolution of the transverse wake function $w$ (depending on the geometry, the dielectric material and the bunch transverse offset with respect to the center of the structure) with the longitudinal charge density distribution of the bunch $n_b(t)$.

\par Thus, when a bunch travels with transverse offset $(x,y)$, parallel to its axis, it drives transverse dipolar wakefields along the bunch described by~\cite{CRAIEVICH:2017}: $ W_{\perp}(t) = w(x,y)\int_0^t n_b(t) dt$, where the bunch front is at $t=0$. 
The amplitude of the wakefields follows the same trend as the running integral of the bunch charge: particles in the back of the bunch are deflected more strongly than those in the front. 
The polarity of $W_{\perp}$ is such that the trailing particles are pulled further towards the dielectric material~\cite{CHAO:1993}.

\par Dielectric capillaries are common tools for generating plasmas in plasma wakefield accelerators (PWFA), as they enable gas injection in a high-vacuum environment~\cite{LEEMANS:2006}. 
The plasma is generated by ionizing the gas injected in the capillary with a high-voltage discharge applied between electrodes at each end~\cite{SPENCE:2000}, or with an ionization field such as a high-intensity laser pulse~\cite{KELDISH:1965} or a relativistic charged particle bunch~\cite{OCONNEL:2006}. 
Capillaries are also employed because they allow shaping the transverse and longitudinal profiles of the plasma density~\cite{LEEMANS:2014}, and locally injecting different gas species for various injection schemes~\cite{CHEN:2012}.

\par In a PWFA based on dielectric capillaries, both plasma and dielectric wakefields can be present at the same time, as the space-charge electric field of the driver bunch may interact with both plasma and dielectric material~\cite{LI:2013,BIAGIONI:2018}. 
In particular, the possible misalignment of the trajectory with respect to the capillary longitudinal axis (e.g., due to jitters or to the finite accuracy of the alignment tools) may induce a deflection to the bunch trajectory.
This could have detrimental effects on the propagation of the bunches and the acceleration quality, by inducing transverse phenomena such as beam-breakup and hosing instability~\cite{WHITTUM:1991,NECHAEVA:2024}.
However, we show in the present work that high-density plasma suppresses the dielectric wakefields driven in the capillary, by screening the space-charge field of the bunch. 
% We discuss the impact on future PWFA, such as EuPRAXIA@SPARC$\_$LAB~\cite{ASSMAN:2020}, where the presence of high-density plasma will suppress the dielectric wakefields driven in a capillary.
 % \section{Experimental Setup}
\par
\begin{figure}[!ht]
\centering
\includegraphics[width=\columnwidth]{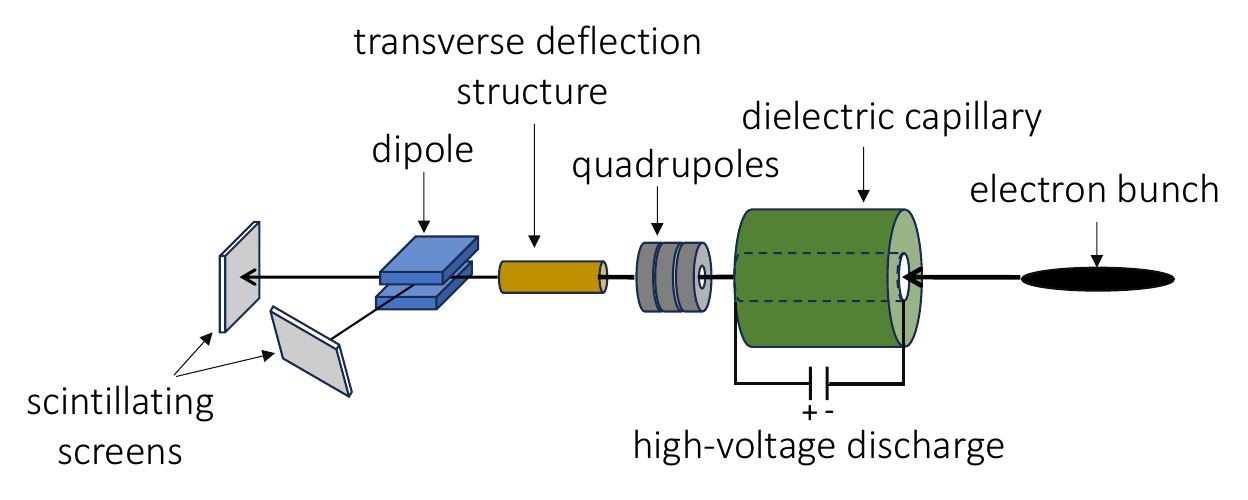}
\caption{Schematic of the SPARC\_LAB experimental setup (not to scale). 
The $e^-$ bunch propagates from right to left.}
\label{fig:1}
\end{figure}

 \textit{Experimental Setup.---}
We performed the measurements at the SPARC$\_$LAB facility~\cite{FERRARIO:2013} (see Fig.~\ref{fig:1}). 
An electron ($e^-$) bunch is generated by illuminating a copper cathode within a radiofrequency (RF) gun with an ultraviolet laser pulse. 
The bunch is accelerated in an RF linac to $E\sim80\,$MeV and focused with transverse root mean square (rms) radius~$\sim 0.4\,$mm at the entrance of a dielectric capillary. 
% A set of electromagnetic quadrupoles transports and images the bunch on a scintillating screen located $d=5.3\,$m downstream of the capillary exit. 
A set of electromagnetic quadrupoles focuses the bunch on a scintillating screen $d=5.3\,$m downstream of the capillary exit for imaging.
A transverse deflection structure (TDS) introduces a head-to-tail vertical correlation to the bunch~\cite{ALESINI:2006}, allowing to obtain time-resolved images of the bunch at the screen position. 
The normalized emittance of the bunch is~$\sim2.4\,$\textmu m (measured with quadrupole scan technique) and rms duration~$\sim1.8\,$ps.
We use a magnetic dipole and an additional scintillating screen to measure the beam energy.  

\par The dielectric capillary has radius $R_c=1\,$mm and length $L=10\,$cm. 
The plasma is generated by a discharge pulse ($\sim455$ A peak current) flowing through the capillary after the introduction of hydrogen gas (${\sim 10\,}$mbar) through a high-speed solenoid valve.  
The capillary is installed in a vacuum chamber connected to the linac with a windowless, three-stage differential pumping system.  
The latter ensures to maintain $10^{-8}\,$mbar in the linac, while flowing the gas, and to preserve the quality of the electron bunch. 
By varying the delay between the bunch arrival time and the peak of the discharge pulse we vary $n_{pe}$, due to recombination of the plasma. 
We measure the average $n_{pe}$ with the Stark broadening technique~\cite{QIAN:2010} between 0.65 and 2.65\,\textmu s after the peak of the discharge pulse and we extrapolate the values of $n_{pe}$ (with~$\sim10\%$ uncertainty) at the delays used in the experiment (between 5.60 and 10.10\,\textmu s) by fitting the measured values with an exponential decay function~\cite{SHPAKOV:2019}.
The plasma electron density is essentially constant along the capillary (see Supplemental Material~\cite{suppl}), but non-uniformities are present at the two extremese because the gas flow expands in the vacuum chamber. 
However, as no capillary material is present, there is also no dielectric wakefield effect along the entrance and exit ramps.
Thus, in the following we neglect these non-uniformities.

\par Current discharge is particularly suited for generating plasma in our experimental conditions because the moderate energy and current of the $e^-$ bunch are not sufficient to field-ionize hydrogen gas. 
Moreover, it is preferable over ionization with a laser pulse because it generates an almost uniform transverse plasma density profile and it requires simpler synchronization and spatial alignemnt methods. 

\par In the following, we maintain the beam trajectory aligned through the center of the quadrupole magnets, and we record images on the screen with the TDS turned on, and the dipole turned off. 
To study the effect of dielectric wakefields and plasma screening, we shift the capillary with a stepper motor in the horizontal direction (perpendicular to the streaking direction of the TDS, see Supplemental Material~\cite{suppl}) and we measure the induced deflection along the bunch on time-resolved images while varying the plasma electron density and, therefore, the plasma skin depth.

% \section{Experimental Results}
\par
\begin{figure}[!h]
\centering
\includegraphics[width=\columnwidth]{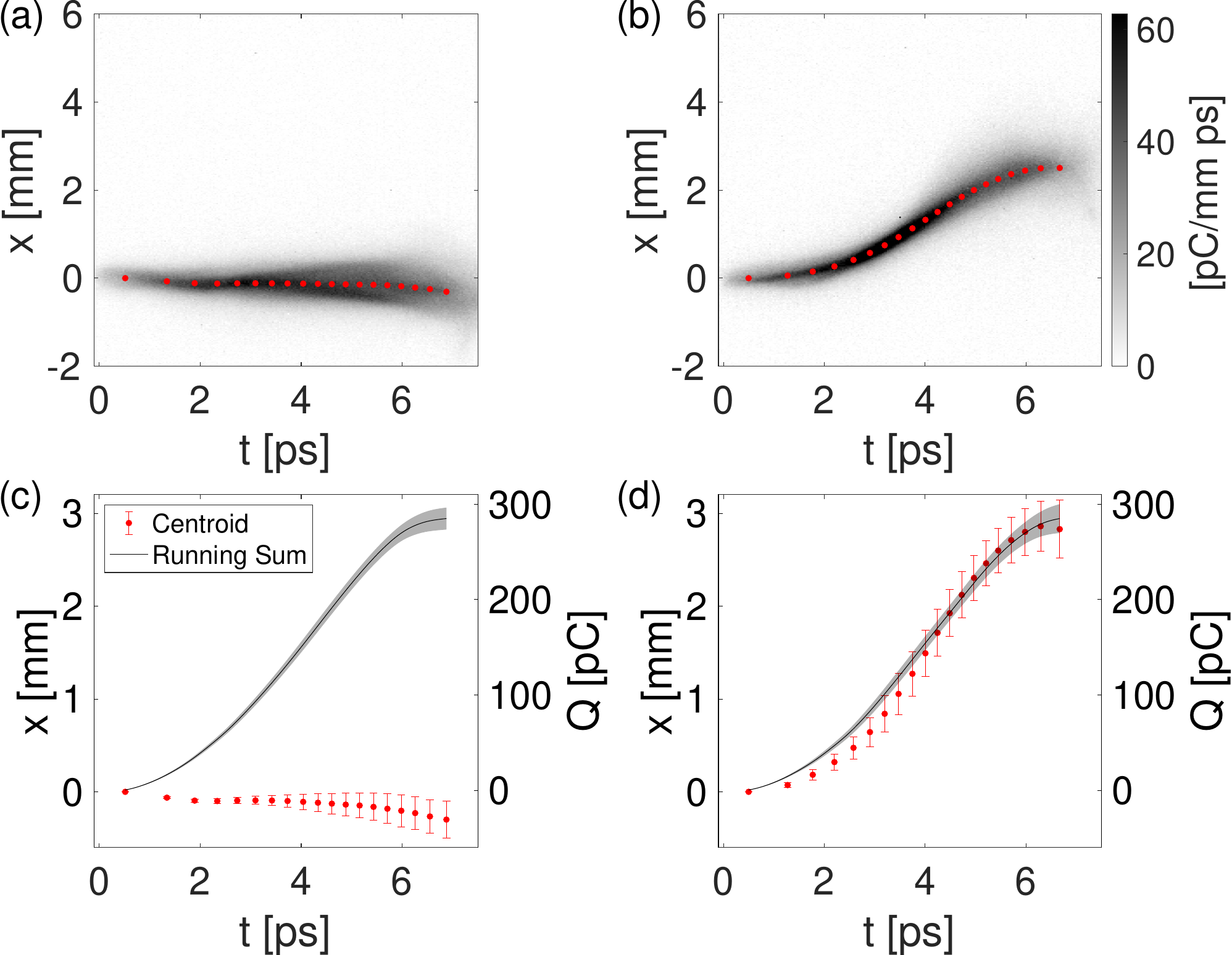}
\caption{ (a,b): single-event, time-resolved images of the 285\,pC $e^-$ bunch for $X=0$ (aligned case) and $X=0.375\,$mm, respectively. 
The bunch propagates from right to left. Red points indicate the centroid position of each longitudinal slice.  
(c,d): average centroid position along the bunch obtained from 100 consecutive events (red points, left-hand side vertical axis; error bars are the standard deviation), corresponding to (a) and (b), respectively. 
Each black line shows the average running sum of the charge along the bunch (right-hand side vertical axis, the shaded area represents the standard deviation). }
\label{fig:2}
\end{figure}

 \textit{Experimental Results.---} 
Figure~\ref{fig:2} shows single-event, time-resolved images of a 285\,pC $e^-$ bunch after travelling along the longitudinal axis of the dielectric capillary (a), and with a horizontal offset $X=0.375\,$mm (b). 
The bunch propagates from right to left (bunch front at $t=0$). 
The red points show the centroid position along the bunch, obtained by fitting the transverse projection of each longitudinal slice with a Gaussian function. 
In the aligned case, the centroid position is essentially constant along the bunch and the variations are much smaller than the bunch transverse size.
% $x\sim 0$ along the entire bunch, because particles are not deflected by the dielectric wakefields in the capillary. 
The average centroid position obtained from 100 consecutive images (red points, left-hand side vertical axis in Fig.~\ref{fig:2}(c), the error bars represent the standard deviation) is not correlated with the running sum of the bunch charge (black line, right-hand side vertical axis, the shaded area is the standard deviation) because $w(0,0)=0$.

\par In the misaligned case (Fig.~\ref{fig:2}(b)), on the contrary, the centroid is clearly deflected, with increasing displacement along the bunch. Figure~\ref{fig:2}(d) shows that the centroid position and the running sum follow the same trend along the bunch. 
This confirms the expectation that the amplitude of the transverse wakefields (and therefore of the transverse deflection) at any time $t$ along the bunch depends on the amount of charge ahead of it, in agreement with the formulation of $W_{\perp}(t)$.

\par The amplitude of the transverse wakefields $W_{\perp}$ reaches a maximum at the back of the bunch. 
We calculate the average wakefield potential experienced by the particles in the last slice of the bunch ($t\sim 6.7\,$ps) as $W_{\perp} = x(t) E /e d L \sim 0.4\,$MV/m. 
We also note that, in both cases, the transverse size slightly increases along the bunch. 
This is due to the fact that the dielectric wakefields also have a quadrupolar (defocusing) component, growing in amplitude along the bunch.

\par
\begin{figure}[!h]
\centering
\includegraphics[width=\columnwidth]{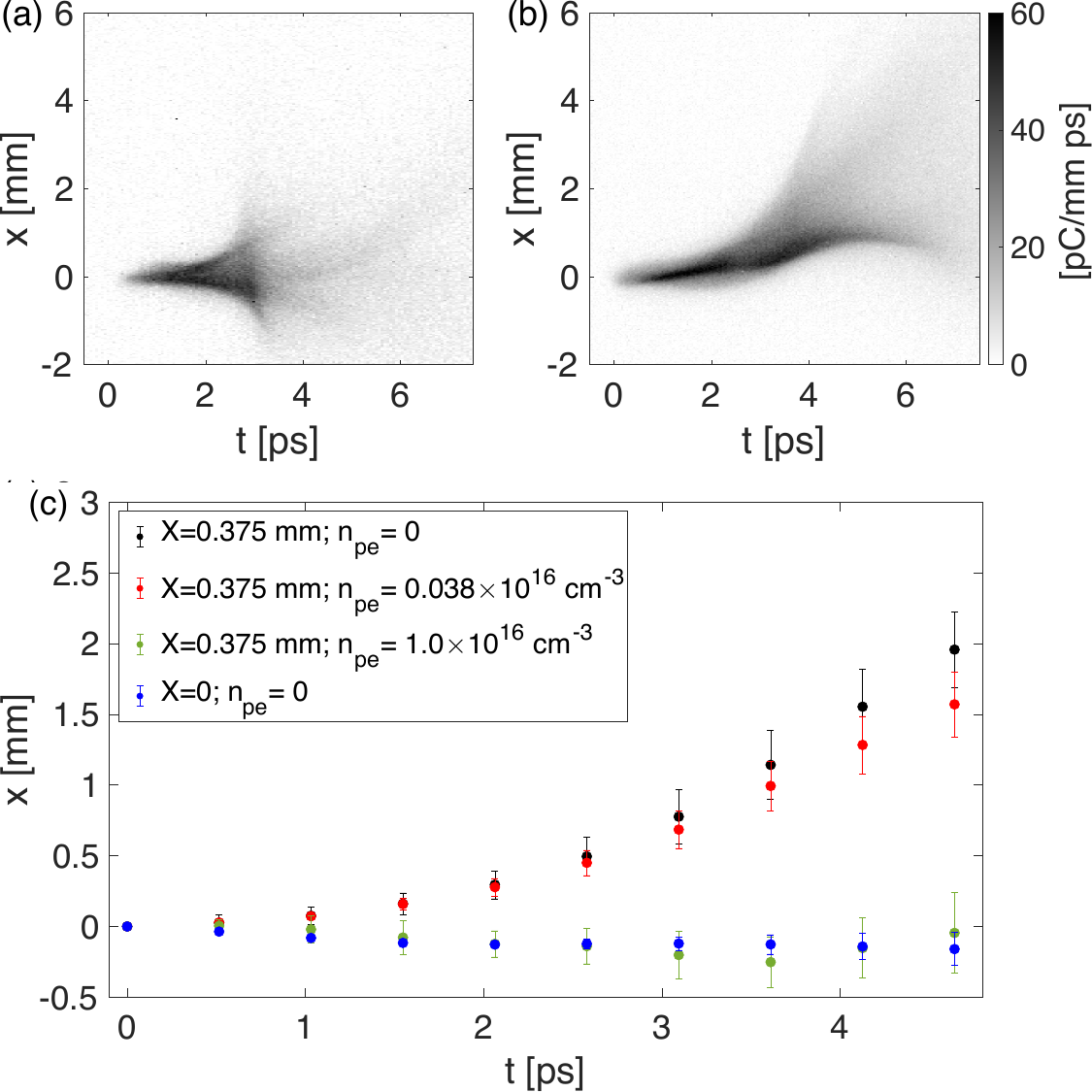}
\caption{ (a,b): single-event, time-resolved images of the $e^-$ bunch with $X=0.375\,$mm (as in Fig.~\ref{fig:2}(b)), and plasma with $n_{pe}=1.0$ and $0.038\cdot 10^{16}\,$cm$^{-3}$, respectively. 
(c): average centroid position along the bunch for $X=0$ and $n_{pe}=0$ (blue points), $X=0.375\,$mm and $n_{pe}=0$ (black points), $n_{pe}=0.038\cdot 10^{16}\,$cm$^{-3}$ (red points) and $n_{pe}=10^{16}\,$cm$^{-3}$ (green points).}
\label{fig:3}
\end{figure}

 When introducing plasma with $n_{pe}=1.0\cdot 10^{16}\,$cm$^{-3}$ while $X=0.375\,$mm (Fig.~\ref{fig:3}(a), same misalignment as Fig.~\ref{fig:2}(b)), the dielectric wakefields are suppressed because the plasma effectively screens the space-charge field of the bunch, since the plasma skin depth $\delta= 0.053\,$mm is much shorter than the distance between bunch and the dielectric surface $R_c-X = 0.625\,$mm. 
Thus, the average position of the centroid along the bunch (green points in (c)) is
% $\sim 0$, 
in agreement with the aligned case with no plasma in the capillary (blue points, same dataset as Fig.~\ref{fig:2}(c)).

\par Slices at $t>3\,$ps have a wider transverse distribution because the bunch also drives plasma wakefields, that are axis-symmetric with respect to the bunch propagation axis.
Additionally, the bunch self-focuses due to the effect of the magnetic field generated by the bunch current itself, that is not compensated by the space-charge force, which is neutralized by the plasma~\cite{CHEN:1987,BAROV:1994,VERRA:2022}. 
As the bunch is progressively focused, it leaves the plasma with increasing divergence along the bunch, and possibly larger emittance, causing the transverse size to increase at the downstream screen. 
Because of this effect, we calculate the position of the centroid only for $t<4.6\,$ps.

\par On the contrary, when $n_{pe}=0.038\cdot 10^{16}\,$cm$^{-3}$ (Fig.~\ref{fig:3}(b)), plasma screening is not effective.
In this case, $\delta = 0.237\,$mm is not short enough to operate full screening, and the space-charge field of the bunch still reaches the dielectric surface, generating dielectric wakefields that affect the trajectory of the trailing part of the bunch. 
Thus, the horizontal position of the centroid along the bunch (red points in Fig.~\ref{fig:3}(c)) varies, in agreement with the case with no plasma (black points, same dataset as Fig.~\ref{fig:2}(d)). 
We note that, in this case, dielectric wakefields (causing the centroid deflection) and plasma wakefields (causing the defocusing effect at the screen) are present at the same time.

% \par We note that, even though the ratio between bunch transverse size and plasma skin depth is larger than unity for all the used values of $n_{pe}$, the occurrence of the current filamentation instability~\cite{LEE:1973,BRET:2010} is not expected because of the low current of the bunch~\cite{ALLEN:2012}.

\par Increasing $n_{pe}$ decreases the distance from the propagation axis beyond which the space-charge field of the bunch is screened. 
Thus, for a small misalignment of the bunch trajectory with respect to the center of the dielectric capillary (i.e., a large distance from the surface), a lower $n_{pe}$ is sufficient to suppress the dielectric wakefields than for a large misalignment (i.e., a short distance from the surface). 

\par
\begin{figure}[h!]
\centering
\includegraphics[width=\columnwidth]{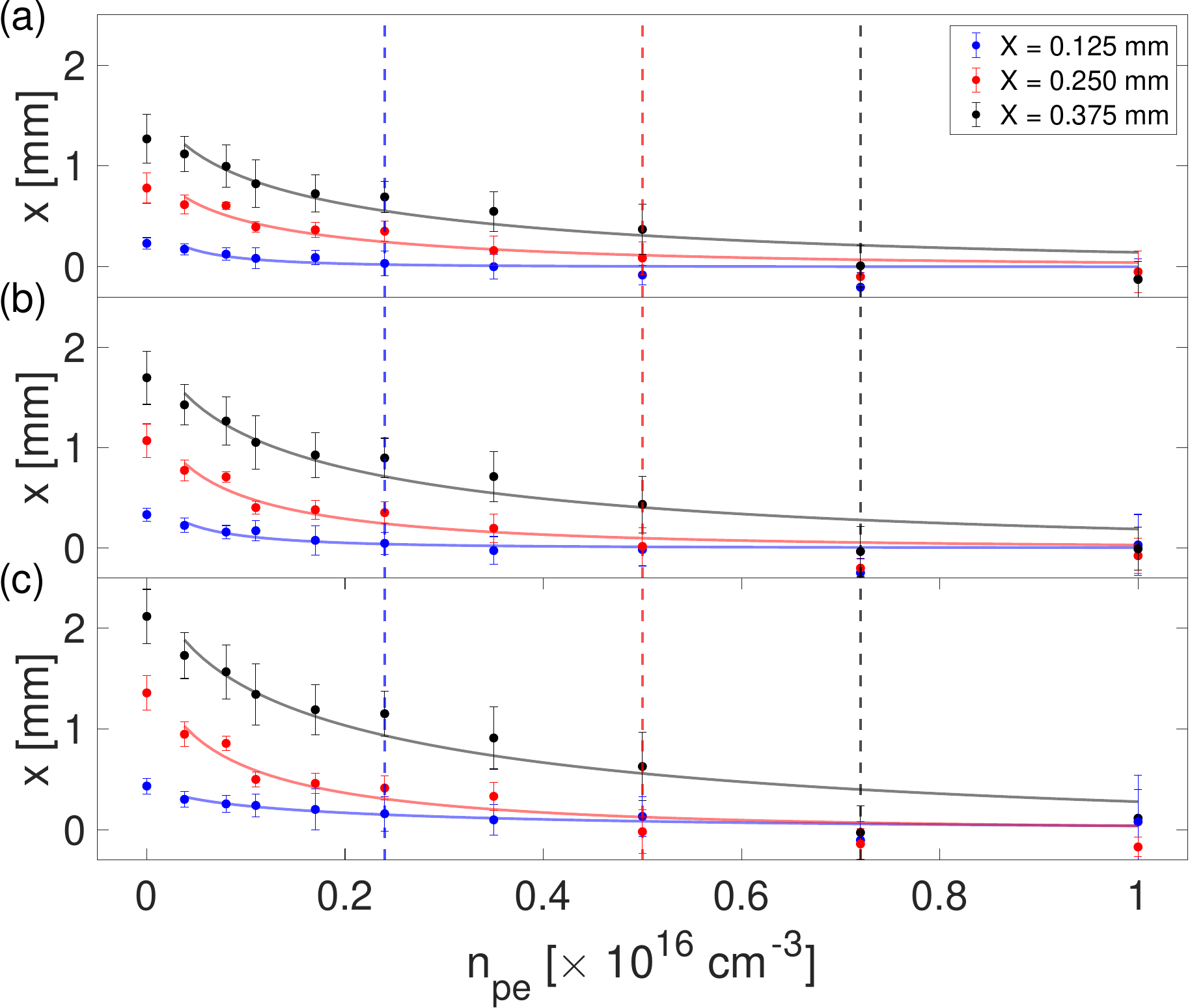}
\caption{
Average centroid position $x$ as a function of $n_{pe}$ for three misalignment distances at longitudinal slices $t=3.6$ ps (a), $4.1$ ps (b), $4.6\,$ps (c) behind the front of the bunch.
Dashed vertical lines indicate the values of $n_{pe}$ after which $x$ is in agreement with the aligned case with no plasma, for each misalignment distance. 
}
\label{fig:4}
\end{figure}

 Figure~\ref{fig:4} shows the position of the centroid at $t=3.6$ ps (a), 4.1 ps (b) and 4.6\, ps (c) behind the front of the bunch as a function of $n_{pe}$, for three misalignment distances (see Legend). 
The displacement, which is due to the effect of the dielectric wakefields, decreases when increasing $n_{pe}$ because the amplitude of the space-charge field reaching the dielectric surface is progressively more screened by plasma. 
The trend is in good agreement with the typical exponential decay expected from plasma screening.
The solid lines show the result of the fit for each dataset, where the distance from the bunch to the dielectric surface is considered as a free parameter. 

\par For $X=0.375\,$mm (black points), full screening (i.e., centroid position in agreement with the aligned case with no plasma) %
% $x\sim0$) 
occurs for $n_{pe}>0.7\cdot10^{16}\,$cm$^{-3}$ (black dashed vertical line), corresponding to $\delta<0.063\,$mm, that is~$\sim 10\,$times shorter than the distance between bunch and capillary surface $R_c-X=0.625\,$mm.
% $\lambda_{pe}<0.39\,$mm$<R_c-X=0.625\,$mm. 
For smaller misalignments, screening occurs at lower $n_{pe}$: for $R_c-X=0.750\,$mm, $n_{pe}>0.5\cdot10^{16}\,$cm$^{-3}$ ($\delta<0.075\,$mm, red points and dashed vertical line); and for $R_c-X=0.875\,$mm, $n_{pe}>0.24\cdot10^{16}\,$cm$^{-3}$ ($\delta<0.109\,$mm, blue points and dashed vertical line).
As expected from the screening process, $\delta$ must be much shorter than the distance between the bunch and dielectric surface to obtain full screening.
We also note that the ratio $(R_c - X)/\delta$ at full screening increases when the misalignment increases.
This is likely due to the finite transverse size of the bunch, as some particles are closer to the dielectric material than those at the bunch center. 
% This is likely due to the fact that some bunch particles are closer to the dielectric material than the center of the bunch, because of the finite transverse size of the bunch.

\par As the bunch travels in plasma, it can drive plasma wakefields that also have a transverse (periodically focusing or defocusing) component~\cite{CHEN:1985, KENIGS:1987}, generally growing along the bunch (see Fig.~\ref{fig:3}(a,b)).
However, the fact that the transition to full screening occurs at the same $n_{pe}$ for all delays along the bunch, as highlighted by the vertical lines in Fig.~\ref{fig:4}, indicates that the screening of the space-charge field is independent of the amplitude of the plasma wakefields.
% (generally growing along the bunch~\cite{CHEN:1985, KENIGS:1987}).
The displacement is larger for larger misalignments, at any $t$ along the bunch where the plasma screening is not fully effective because $w(x,y)$ increases. 

\par The experimental evidence we present in this work will have significant implications for designing and operating plasma-based accelerators.
In particular, the direct observation of space-charge screening in plasma could be employed for managing misalignments and optimizing plasma conditions to minimize undesired wakefield effect.

\par Considering future PWFA, the operational $n_{pe}$ and the transverse size of the capillary must be chosen so that the dielectric wakefields induced by misalignments (e.g., caused by jitters and finite accuracy of the alignment tools) are effectively screened by the plasma. 
In the case of EuPRAXIA@SPARC\_LAB~\cite{ASSMAN:2020}, the diameter of the dielectric capillary will be 2\,mm and the operational density $n_{pe}=10^{16}\,$cm$^{-3}$.
The maximum misalignment allowed to avoid the dielectric wakefields effect would therefore be even larger than $X=0.375\,$mm, as shown in this study, which is much larger than the state-of-the-art capabilities to align and control the trajectory of relativistic particle bunches (e.g., better than 20\,\textmu m in our experiment).
The measurement we present in this work could also be extended to a plasma electron density diagnostics, by measuring the maximum misalignment for which full screening of the space-charge field occurs.

% \section{Summary}
\par \textit{Summary.---}
We investigated, with a non-invasive measurement, one of the fundamental characteristics of plasma, that is the screening of electromagnetic fields. 
We showed, using a plasma wakefield accelerator based on a dielectric capillary, that plasma screens the space-charge field of a relativistic $e^-$ bunch by measuring the suppression of dielectric wakefields.
The effect of transverse dielectric wakefields increases along the bunch, following the same trend as the running sum of the bunch charge, in agreement with the theory. 
We observed a progressively increasing screening effect when increasing the plasma electron density, and therefore decreasing the plasma skin depth. 
We measured full screening to occur along the entire bunch when the distance from the bunch propagation axis to the dielectric surface is more than 10 times longer than the plasma skin depth. 
When full screening is not achieved, dielectric and plasma wakefields are present simultaneously.
We also discussed the implications for future plasma-based accelerators based on dielectric capillaries.

\begin{acknowledgments}
This work has been partially supported by the European Commission in the Seventh Framework Programme, grant agreement 312453-EuCARD-2, the European Union Horizon 2020 research and innovation program, grant agreement no. 653782 (EuPRAXIA) and the INFN with the GRANT73/PLADIP and SL COMB2FEL grants. 
We thank G.~Grilli and T.~De Nardis for the development of the HV discharge pulser, F.~Anelli for the technical support and M.~Zottola for the experimental chamber installation. 
\end{acknowledgments}

\bibliography{resubmission_2}% Produces the bibliography via BibTeX.

\begin{thebibliography}{38}%
\makeatletter
\providecommand \@ifxundefined [1]{%
 \@ifx{#1\undefined}
}%
\providecommand \@ifnum [1]{%
 \ifnum #1\expandafter \@firstoftwo
 \else \expandafter \@secondoftwo
 \fi
}%
\providecommand \@ifx [1]{%
 \ifx #1\expandafter \@firstoftwo
 \else \expandafter \@secondoftwo
 \fi
}%
\providecommand \natexlab [1]{#1}%
\providecommand \enquote  [1]{``#1''}%
\providecommand \bibnamefont  [1]{#1}%
\providecommand \bibfnamefont [1]{#1}%
\providecommand \citenamefont [1]{#1}%
\providecommand \href@noop [0]{\@secondoftwo}%
\providecommand \href [0]{\begingroup \@sanitize@url \@href}%
\providecommand \@href[1]{\@@startlink{#1}\@@href}%
\providecommand \@@href[1]{\endgroup#1\@@endlink}%
\providecommand \@sanitize@url [0]{\catcode `\\12\catcode `\$12\catcode
  `\&12\catcode `\#12\catcode `\^12\catcode `\_12\catcode `\%12\relax}%
\providecommand \@@startlink[1]{}%
\providecommand \@@endlink[0]{}%
\providecommand \url  [0]{\begingroup\@sanitize@url \@url }%
\providecommand \@url [1]{\endgroup\@href {#1}{\urlprefix }}%
\providecommand \urlprefix  [0]{URL }%
\providecommand \Eprint [0]{\href }%
\providecommand \doibase [0]{https://doi.org/}%
\providecommand \selectlanguage [0]{\@gobble}%
\providecommand \bibinfo  [0]{\@secondoftwo}%
\providecommand \bibfield  [0]{\@secondoftwo}%
\providecommand \translation [1]{[#1]}%
\providecommand \BibitemOpen [0]{}%
\providecommand \bibitemStop [0]{}%
\providecommand \bibitemNoStop [0]{.\EOS\space}%
\providecommand \EOS [0]{\spacefactor3000\relax}%
\providecommand \BibitemShut  [1]{\csname bibitem#1\endcsname}%
\let\auto@bib@innerbib\@empty
%</preamble>
\bibitem [{\citenamefont {Chen}(2016)}]{CHEN:BOOK}%
  \BibitemOpen
  \bibfield  {author} {\bibinfo {author} {\bibfnamefont {F.~F.}\ \bibnamefont
  {Chen}},\ }\href {https://doi.org/10.1007/978-3-319-22309-4_1} {\emph
  {\bibinfo {title} {Introduction to Plasma Physics and Controlled Fusion}}}\
  (\bibinfo  {publisher} {Springer International Publishing},\ \bibinfo {year}
  {2016})\BibitemShut {NoStop}%
\bibitem [{\citenamefont {Otsuka}\ \emph {et~al.}(2015)\citenamefont {Otsuka},
  \citenamefont {Hada}, \citenamefont {Shinohara},\ and\ \citenamefont
  {Tanikawa}}]{OSTUKA:2015}%
  \BibitemOpen
  \bibfield  {author} {\bibinfo {author} {\bibfnamefont {F.}~\bibnamefont
  {Otsuka}}, \bibinfo {author} {\bibfnamefont {T.}~\bibnamefont {Hada}},
  \bibinfo {author} {\bibfnamefont {S.}~\bibnamefont {Shinohara}},\ and\
  \bibinfo {author} {\bibfnamefont {T.}~\bibnamefont {Tanikawa}},\ }\bibfield
  {title} {\bibinfo {title} {Penetration of a radio frequency electromagnetic
  field into a magnetized plasma: one-dimensional pic simulation studies},\
  }\href {https://doi.org/10.1186/s40623-015-0244-9} {\bibfield  {journal}
  {\bibinfo  {journal} {Earth, Planets and Space}\ }\textbf {\bibinfo {volume}
  {67}},\ \bibinfo {pages} {85} (\bibinfo {year} {2015})}\BibitemShut {NoStop}%
\bibitem [{\citenamefont {Shinohara}\ and\ \citenamefont
  {Shamrai}(2002)}]{SHINOHARA:2002}%
  \BibitemOpen
  \bibfield  {author} {\bibinfo {author} {\bibfnamefont {S.}~\bibnamefont
  {Shinohara}}\ and\ \bibinfo {author} {\bibfnamefont {K.~P.}\ \bibnamefont
  {Shamrai}},\ }\bibfield  {title} {\bibinfo {title} {Effect of electrostatic
  waves on a rf field penetration into highly collisional helicon plasmas},\
  }\href {https://doi.org/https://doi.org/10.1016/S0040-6090(02)00041-X}
  {\bibfield  {journal} {\bibinfo  {journal} {Thin Solid Films}\ }\textbf
  {\bibinfo {volume} {407}},\ \bibinfo {pages} {215} (\bibinfo {year}
  {2002})},\ \bibinfo {note} {proceedings of the 14th Symposium on Plasma
  Science for Materials}\BibitemShut {NoStop}%
\bibitem [{\citenamefont {Shoji}(1980)}]{SHOJI:1980}%
  \BibitemOpen
  \bibfield  {author} {\bibinfo {author} {\bibfnamefont {T.}~\bibnamefont
  {Shoji}},\ }\bibfield  {title} {\bibinfo {title} {Description of
  radio-frequency plugging and heating in terms of plasma impedance},\ }\href
  {https://doi.org/10.1143/JPSJ.49.327} {\bibfield  {journal} {\bibinfo
  {journal} {Journal of the Physical Society of Japan}\ }\textbf {\bibinfo
  {volume} {49}},\ \bibinfo {pages} {327} (\bibinfo {year} {1980})},\ \Eprint
  {https://arxiv.org/abs/https://doi.org/10.1143/JPSJ.49.327}
  {https://doi.org/10.1143/JPSJ.49.327} \BibitemShut {NoStop}%
\bibitem [{\citenamefont {Whelan}\ and\ \citenamefont
  {Stenzel}(1981)}]{WHELAN:1981}%
  \BibitemOpen
  \bibfield  {author} {\bibinfo {author} {\bibfnamefont {D.~A.}\ \bibnamefont
  {Whelan}}\ and\ \bibinfo {author} {\bibfnamefont {R.~L.}\ \bibnamefont
  {Stenzel}},\ }\bibfield  {title} {\bibinfo {title} {Electromagnetic-wave
  excitation in a large laboratory beam-plasma system},\ }\href
  {https://doi.org/10.1103/PhysRevLett.47.95} {\bibfield  {journal} {\bibinfo
  {journal} {Phys. Rev. Lett.}\ }\textbf {\bibinfo {volume} {47}},\ \bibinfo
  {pages} {95} (\bibinfo {year} {1981})}\BibitemShut {NoStop}%
\bibitem [{\citenamefont {Chen}\ \emph {et~al.}(1985)\citenamefont {Chen},
  \citenamefont {Dawson}, \citenamefont {Huff},\ and\ \citenamefont
  {Katsouleas}}]{CHEN:1985}%
  \BibitemOpen
  \bibfield  {author} {\bibinfo {author} {\bibfnamefont {P.}~\bibnamefont
  {Chen}}, \bibinfo {author} {\bibfnamefont {J.~M.}\ \bibnamefont {Dawson}},
  \bibinfo {author} {\bibfnamefont {R.~W.}\ \bibnamefont {Huff}},\ and\
  \bibinfo {author} {\bibfnamefont {T.}~\bibnamefont {Katsouleas}},\ }\bibfield
   {title} {\bibinfo {title} {Acceleration of electrons by the interaction of a
  bunched electron beam with a plasma},\ }\href
  {https://doi.org/10.1103/PhysRevLett.54.693} {\bibfield  {journal} {\bibinfo
  {journal} {Physical Review Letters}\ }\textbf {\bibinfo {volume} {54}},\
  \bibinfo {pages} {693} (\bibinfo {year} {1985})}\BibitemShut {NoStop}%
\bibitem [{\citenamefont {Keinigs}\ and\ \citenamefont
  {Jones}(1987)}]{KENIGS:1987}%
  \BibitemOpen
  \bibfield  {author} {\bibinfo {author} {\bibfnamefont {R.}~\bibnamefont
  {Keinigs}}\ and\ \bibinfo {author} {\bibfnamefont {M.~E.}\ \bibnamefont
  {Jones}},\ }\bibfield  {title} {\bibinfo {title} {Two‐dimensional dynamics
  of the plasma wakefield accelerator},\ }\href
  {https://doi.org/10.1063/1.866183} {\bibfield  {journal} {\bibinfo  {journal}
  {The Physics of Fluids}\ }\textbf {\bibinfo {volume} {30}},\ \bibinfo {pages}
  {252} (\bibinfo {year} {1987})}\BibitemShut {NoStop}%
\bibitem [{\citenamefont {Jackson}(1962)}]{JACKSON:BOOK}%
  \BibitemOpen
  \bibfield  {author} {\bibinfo {author} {\bibfnamefont {J.~D.}\ \bibnamefont
  {Jackson}},\ }\href@noop {} {\emph {\bibinfo {title} {Classical
  Electrodynamics}}}\ (\bibinfo  {publisher} {John Wiley \& Sons},\ \bibinfo
  {year} {1962})\BibitemShut {NoStop}%
\bibitem [{\citenamefont {Bane}\ \emph {et~al.}(1985)\citenamefont {Bane},
  \citenamefont {Wilson},\ and\ \citenamefont {Weiland}}]{BANE:1984}%
  \BibitemOpen
  \bibfield  {author} {\bibinfo {author} {\bibfnamefont {K.~L.}\ \bibnamefont
  {Bane}}, \bibinfo {author} {\bibfnamefont {P.~B.}\ \bibnamefont {Wilson}},\
  and\ \bibinfo {author} {\bibfnamefont {T.}~\bibnamefont {Weiland}},\
  }\bibfield  {title} {\bibinfo {title} {{Wake Fields and Wake Field
  Acceleration}},\ }\href {https://doi.org/10.1063/1.35182} {\bibfield
  {journal} {\bibinfo  {journal} {AIP Conf. Proc.}\ }\textbf {\bibinfo {volume}
  {127}},\ \bibinfo {pages} {875} (\bibinfo {year} {1985})}\BibitemShut
  {NoStop}%
\bibitem [{\citenamefont {Gai}\ \emph {et~al.}(1988)\citenamefont {Gai},
  \citenamefont {Schoessow}, \citenamefont {Cole}, \citenamefont {Konecny},
  \citenamefont {Norem}, \citenamefont {Rosenzweig},\ and\ \citenamefont
  {Simpson}}]{GAI:1988}%
  \BibitemOpen
  \bibfield  {author} {\bibinfo {author} {\bibfnamefont {W.}~\bibnamefont
  {Gai}}, \bibinfo {author} {\bibfnamefont {P.}~\bibnamefont {Schoessow}},
  \bibinfo {author} {\bibfnamefont {B.}~\bibnamefont {Cole}}, \bibinfo {author}
  {\bibfnamefont {R.}~\bibnamefont {Konecny}}, \bibinfo {author} {\bibfnamefont
  {J.}~\bibnamefont {Norem}}, \bibinfo {author} {\bibfnamefont
  {J.}~\bibnamefont {Rosenzweig}},\ and\ \bibinfo {author} {\bibfnamefont
  {J.}~\bibnamefont {Simpson}},\ }\bibfield  {title} {\bibinfo {title}
  {Experimental demonstration of wake-field effects in dielectric structures},\
  }\href {https://doi.org/10.1103/PhysRevLett.61.2756} {\bibfield  {journal}
  {\bibinfo  {journal} {Physical Review Letters}\ }\textbf {\bibinfo {volume}
  {61}},\ \bibinfo {pages} {2756} (\bibinfo {year} {1988})}\BibitemShut
  {NoStop}%
\bibitem [{\citenamefont {Keinigs}\ \emph {et~al.}(1989)\citenamefont
  {Keinigs}, \citenamefont {Peter},\ and\ \citenamefont
  {Jones}}]{KEINIGS:1989}%
  \BibitemOpen
  \bibfield  {author} {\bibinfo {author} {\bibfnamefont {R.}~\bibnamefont
  {Keinigs}}, \bibinfo {author} {\bibfnamefont {W.}~\bibnamefont {Peter}},\
  and\ \bibinfo {author} {\bibfnamefont {M.~E.}\ \bibnamefont {Jones}},\
  }\bibfield  {title} {\bibinfo {title} {{A comparison of the dielectric and
  plasma wakefield accelerators}},\ }\href {https://doi.org/10.1063/1.858920}
  {\bibfield  {journal} {\bibinfo  {journal} {Physics of Fluids B: Plasma
  Physics}\ }\textbf {\bibinfo {volume} {1}},\ \bibinfo {pages} {1872}
  (\bibinfo {year} {1989})},\ \Eprint
  {https://arxiv.org/abs/https://pubs.aip.org/aip/pfb/article-pdf/1/9/1872/12643661/1872\_1\_online.pdf}
  {https://pubs.aip.org/aip/pfb/article-pdf/1/9/1872/12643661/1872\_1\_online.pdf}
  \BibitemShut {NoStop}%
\bibitem [{\citenamefont {Ng}(1990)}]{NG:1990}%
  \BibitemOpen
  \bibfield  {author} {\bibinfo {author} {\bibfnamefont {K.-Y.}\ \bibnamefont
  {Ng}},\ }\bibfield  {title} {\bibinfo {title} {Wake fields in a
  dielectric-lined waveguide},\ }\href
  {https://doi.org/10.1103/PhysRevD.42.1819} {\bibfield  {journal} {\bibinfo
  {journal} {Phys. Rev. D}\ }\textbf {\bibinfo {volume} {42}},\ \bibinfo
  {pages} {1819} (\bibinfo {year} {1990})}\BibitemShut {NoStop}%
\bibitem [{\citenamefont {Park}\ and\ \citenamefont
  {Hirshfield}(2000)}]{PARK:2000}%
  \BibitemOpen
  \bibfield  {author} {\bibinfo {author} {\bibfnamefont {S.~Y.}\ \bibnamefont
  {Park}}\ and\ \bibinfo {author} {\bibfnamefont {J.~L.}\ \bibnamefont
  {Hirshfield}},\ }\bibfield  {title} {\bibinfo {title} {Theory of wakefields
  in a dielectric-lined waveguide},\ }\href
  {https://doi.org/10.1103/PhysRevE.62.1266} {\bibfield  {journal} {\bibinfo
  {journal} {Physical Review E}\ }\textbf {\bibinfo {volume} {62}},\ \bibinfo
  {pages} {1266} (\bibinfo {year} {2000})}\BibitemShut {NoStop}%
\bibitem [{\citenamefont {Baturin}\ and\ \citenamefont
  {Kanareykin}(2014)}]{BATURIN:2014}%
  \BibitemOpen
  \bibfield  {author} {\bibinfo {author} {\bibfnamefont {S.~S.}\ \bibnamefont
  {Baturin}}\ and\ \bibinfo {author} {\bibfnamefont {A.~D.}\ \bibnamefont
  {Kanareykin}},\ }\bibfield  {title} {\bibinfo {title} {Cherenkov radiation
  from short relativistic bunches: General approach},\ }\href
  {https://doi.org/10.1103/PhysRevLett.113.214801} {\bibfield  {journal}
  {\bibinfo  {journal} {Physical Review Letters}\ }\textbf {\bibinfo {volume}
  {113}},\ \bibinfo {pages} {214801} (\bibinfo {year} {2014})}\BibitemShut
  {NoStop}%
\bibitem [{\citenamefont {Bettoni}\ \emph {et~al.}(2016)\citenamefont
  {Bettoni}, \citenamefont {Craievich}, \citenamefont {Lutman},\ and\
  \citenamefont {Pedrozzi}}]{BETTONI:2016}%
  \BibitemOpen
  \bibfield  {author} {\bibinfo {author} {\bibfnamefont {S.}~\bibnamefont
  {Bettoni}}, \bibinfo {author} {\bibfnamefont {P.}~\bibnamefont {Craievich}},
  \bibinfo {author} {\bibfnamefont {A.~A.}\ \bibnamefont {Lutman}},\ and\
  \bibinfo {author} {\bibfnamefont {M.}~\bibnamefont {Pedrozzi}},\ }\bibfield
  {title} {\bibinfo {title} {Temporal profile measurements of relativistic
  electron bunch based on wakefield generation},\ }\href
  {https://doi.org/10.1103/PhysRevAccelBeams.19.021304} {\bibfield  {journal}
  {\bibinfo  {journal} {Physical Review Accelerators and Beams}\ }\textbf
  {\bibinfo {volume} {19}},\ \bibinfo {pages} {021304} (\bibinfo {year}
  {2016})}\BibitemShut {NoStop}%
\bibitem [{\citenamefont {Craievich}\ and\ \citenamefont
  {Lutman}(2017)}]{CRAIEVICH:2017}%
  \BibitemOpen
  \bibfield  {author} {\bibinfo {author} {\bibfnamefont {P.}~\bibnamefont
  {Craievich}}\ and\ \bibinfo {author} {\bibfnamefont {A.~A.}\ \bibnamefont
  {Lutman}},\ }\bibfield  {title} {\bibinfo {title} {Effects of the quadrupole
  wakefields in a passive streaker},\ }\href
  {https://doi.org/https://doi.org/10.1016/j.nima.2016.10.010} {\bibfield
  {journal} {\bibinfo  {journal} {Nuclear Instruments and Methods in Physics
  Research Section A: Accelerators, Spectrometers, Detectors and Associated
  Equipment}\ }\textbf {\bibinfo {volume} {865}},\ \bibinfo {pages} {55}
  (\bibinfo {year} {2017})}\BibitemShut {NoStop}%
\bibitem [{\citenamefont {O'Shea}\ \emph {et~al.}(2020)\citenamefont {O'Shea},
  \citenamefont {Andonian}, \citenamefont {Baturin}, \citenamefont {Clarke},
  \citenamefont {Hoang}, \citenamefont {Hogan}, \citenamefont {Naranjo},
  \citenamefont {Williams}, \citenamefont {Yakimenko},\ and\ \citenamefont
  {Rosenzweig}}]{OSHEA:2020}%
  \BibitemOpen
  \bibfield  {author} {\bibinfo {author} {\bibfnamefont {B.~D.}\ \bibnamefont
  {O'Shea}}, \bibinfo {author} {\bibfnamefont {G.}~\bibnamefont {Andonian}},
  \bibinfo {author} {\bibfnamefont {S.~S.}\ \bibnamefont {Baturin}}, \bibinfo
  {author} {\bibfnamefont {C.~I.}\ \bibnamefont {Clarke}}, \bibinfo {author}
  {\bibfnamefont {P.~D.}\ \bibnamefont {Hoang}}, \bibinfo {author}
  {\bibfnamefont {M.~J.}\ \bibnamefont {Hogan}}, \bibinfo {author}
  {\bibfnamefont {B.}~\bibnamefont {Naranjo}}, \bibinfo {author} {\bibfnamefont
  {O.~B.}\ \bibnamefont {Williams}}, \bibinfo {author} {\bibfnamefont
  {V.}~\bibnamefont {Yakimenko}},\ and\ \bibinfo {author} {\bibfnamefont
  {J.~B.}\ \bibnamefont {Rosenzweig}},\ }\bibfield  {title} {\bibinfo {title}
  {Suppression of deflecting forces in planar-symmetric dielectric wakefield
  accelerating structures with elliptical bunches},\ }\href
  {https://doi.org/10.1103/PhysRevLett.124.104801} {\bibfield  {journal}
  {\bibinfo  {journal} {Physical Review Letters}\ }\textbf {\bibinfo {volume}
  {124}},\ \bibinfo {pages} {104801} (\bibinfo {year} {2020})}\BibitemShut
  {NoStop}%
\bibitem [{\citenamefont {Saveliev}\ \emph {et~al.}(2022)\citenamefont
  {Saveliev}, \citenamefont {Overton}, \citenamefont {Pacey}, \citenamefont
  {Joshi}, \citenamefont {Mathisen}, \citenamefont {Muratori}, \citenamefont
  {Thompson}, \citenamefont {King},\ and\ \citenamefont {Xia}}]{SAVELIEV:2022}%
  \BibitemOpen
  \bibfield  {author} {\bibinfo {author} {\bibfnamefont {Y.}~\bibnamefont
  {Saveliev}}, \bibinfo {author} {\bibfnamefont {T.~J.}\ \bibnamefont
  {Overton}}, \bibinfo {author} {\bibfnamefont {T.~H.}\ \bibnamefont {Pacey}},
  \bibinfo {author} {\bibfnamefont {N.}~\bibnamefont {Joshi}}, \bibinfo
  {author} {\bibfnamefont {S.}~\bibnamefont {Mathisen}}, \bibinfo {author}
  {\bibfnamefont {B.~D.}\ \bibnamefont {Muratori}}, \bibinfo {author}
  {\bibfnamefont {N.}~\bibnamefont {Thompson}}, \bibinfo {author}
  {\bibfnamefont {M.~P.}\ \bibnamefont {King}},\ and\ \bibinfo {author}
  {\bibfnamefont {G.}~\bibnamefont {Xia}},\ }\bibfield  {title} {\bibinfo
  {title} {Experimental study of transverse effects in planar dielectric
  wakefield accelerating structures with elliptical beams},\ }\href
  {https://doi.org/10.1103/PhysRevAccelBeams.25.081302} {\bibfield  {journal}
  {\bibinfo  {journal} {Physical Review Accelerators and Beams}\ }\textbf
  {\bibinfo {volume} {25}},\ \bibinfo {pages} {081302} (\bibinfo {year}
  {2022})}\BibitemShut {NoStop}%
\bibitem [{\citenamefont {Chao}(1993)}]{CHAO:1993}%
  \BibitemOpen
  \bibfield  {author} {\bibinfo {author} {\bibfnamefont {A.~W.}\ \bibnamefont
  {Chao}},\ }\href@noop {} {\emph {\bibinfo {title} {{Physics of collective
  beam instabilities in high-energy accelerators}}}}\ (\bibinfo  {publisher}
  {John Wiley \& Sons, Inc.},\ \bibinfo {year} {1993})\BibitemShut {NoStop}%
\bibitem [{\citenamefont {Leemans}\ \emph {et~al.}(2006)\citenamefont
  {Leemans}, \citenamefont {Nagler}, \citenamefont {Gonsalves}, \citenamefont
  {T{\'o}th}, \citenamefont {Nakamura}, \citenamefont {Geddes}, \citenamefont
  {Esarey}, \citenamefont {Schroeder},\ and\ \citenamefont
  {Hooker}}]{LEEMANS:2006}%
  \BibitemOpen
  \bibfield  {author} {\bibinfo {author} {\bibfnamefont {W.~P.}\ \bibnamefont
  {Leemans}}, \bibinfo {author} {\bibfnamefont {B.}~\bibnamefont {Nagler}},
  \bibinfo {author} {\bibfnamefont {A.~J.}\ \bibnamefont {Gonsalves}}, \bibinfo
  {author} {\bibfnamefont {C.}~\bibnamefont {T{\'o}th}}, \bibinfo {author}
  {\bibfnamefont {K.}~\bibnamefont {Nakamura}}, \bibinfo {author}
  {\bibfnamefont {C.~G.~R.}\ \bibnamefont {Geddes}}, \bibinfo {author}
  {\bibfnamefont {E.}~\bibnamefont {Esarey}}, \bibinfo {author} {\bibfnamefont
  {C.~B.}\ \bibnamefont {Schroeder}},\ and\ \bibinfo {author} {\bibfnamefont
  {S.~M.}\ \bibnamefont {Hooker}},\ }\bibfield  {title} {\bibinfo {title} {Gev
  electron beams from a centimetre-scale accelerator},\ }\href
  {https://doi.org/10.1038/nphys418} {\bibfield  {journal} {\bibinfo  {journal}
  {Nature Physics}\ }\textbf {\bibinfo {volume} {2}},\ \bibinfo {pages} {696}
  (\bibinfo {year} {2006})}\BibitemShut {NoStop}%
\bibitem [{\citenamefont {Spence}\ and\ \citenamefont
  {Hooker}(2000)}]{SPENCE:2000}%
  \BibitemOpen
  \bibfield  {author} {\bibinfo {author} {\bibfnamefont {D.~J.}\ \bibnamefont
  {Spence}}\ and\ \bibinfo {author} {\bibfnamefont {S.~M.}\ \bibnamefont
  {Hooker}},\ }\bibfield  {title} {\bibinfo {title} {Investigation of a
  hydrogen plasma waveguide},\ }\href
  {https://doi.org/10.1103/PhysRevE.63.015401} {\bibfield  {journal} {\bibinfo
  {journal} {Phys. Rev. E}\ }\textbf {\bibinfo {volume} {63}},\ \bibinfo
  {pages} {015401} (\bibinfo {year} {2000})}\BibitemShut {NoStop}%
\bibitem [{\citenamefont {Keldysh}(1965)}]{KELDISH:1965}%
  \BibitemOpen
  \bibfield  {author} {\bibinfo {author} {\bibfnamefont {L.~V.}\ \bibnamefont
  {Keldysh}},\ }\bibfield  {title} {\bibinfo {title} {{Ionization in the Field
  of a Strong Electromagnetic Wave}},\ }\href@noop {} {\bibfield  {journal}
  {\bibinfo  {journal} {J. Exp. Theor. Phys.}\ }\textbf {\bibinfo {volume}
  {20}},\ \bibinfo {pages} {1307} (\bibinfo {year} {1965})}\BibitemShut
  {NoStop}%
\bibitem [{\citenamefont {O'Connell}\ \emph {et~al.}(2006)\citenamefont
  {O'Connell}, \citenamefont {Barnes}, \citenamefont {Decker}, \citenamefont
  {Hogan}, \citenamefont {Iverson}, \citenamefont {Krejcik}, \citenamefont
  {Siemann}, \citenamefont {Walz}, \citenamefont {Clayton}, \citenamefont
  {Huang}, \citenamefont {Johnson}, \citenamefont {Joshi}, \citenamefont {Lu},
  \citenamefont {Marsh}, \citenamefont {Mori}, \citenamefont {Zhou},
  \citenamefont {Deng}, \citenamefont {Katsouleas}, \citenamefont {Muggli},\
  and\ \citenamefont {Oz}}]{OCONNEL:2006}%
  \BibitemOpen
  \bibfield  {author} {\bibinfo {author} {\bibfnamefont {C.~L.}\ \bibnamefont
  {O'Connell}}, \bibinfo {author} {\bibfnamefont {C.~D.}\ \bibnamefont
  {Barnes}}, \bibinfo {author} {\bibfnamefont {F.-J.}\ \bibnamefont {Decker}},
  \bibinfo {author} {\bibfnamefont {M.~J.}\ \bibnamefont {Hogan}}, \bibinfo
  {author} {\bibfnamefont {R.}~\bibnamefont {Iverson}}, \bibinfo {author}
  {\bibfnamefont {P.}~\bibnamefont {Krejcik}}, \bibinfo {author} {\bibfnamefont
  {R.}~\bibnamefont {Siemann}}, \bibinfo {author} {\bibfnamefont {D.~R.}\
  \bibnamefont {Walz}}, \bibinfo {author} {\bibfnamefont {C.~E.}\ \bibnamefont
  {Clayton}}, \bibinfo {author} {\bibfnamefont {C.}~\bibnamefont {Huang}},
  \bibinfo {author} {\bibfnamefont {D.~K.}\ \bibnamefont {Johnson}}, \bibinfo
  {author} {\bibfnamefont {C.}~\bibnamefont {Joshi}}, \bibinfo {author}
  {\bibfnamefont {W.}~\bibnamefont {Lu}}, \bibinfo {author} {\bibfnamefont
  {K.~A.}\ \bibnamefont {Marsh}}, \bibinfo {author} {\bibfnamefont
  {W.}~\bibnamefont {Mori}}, \bibinfo {author} {\bibfnamefont {M.}~\bibnamefont
  {Zhou}}, \bibinfo {author} {\bibfnamefont {S.}~\bibnamefont {Deng}}, \bibinfo
  {author} {\bibfnamefont {T.}~\bibnamefont {Katsouleas}}, \bibinfo {author}
  {\bibfnamefont {P.}~\bibnamefont {Muggli}},\ and\ \bibinfo {author}
  {\bibfnamefont {E.}~\bibnamefont {Oz}},\ }\bibfield  {title} {\bibinfo
  {title} {Plasma production via field ionization},\ }\href
  {https://doi.org/10.1103/PhysRevSTAB.9.101301} {\bibfield  {journal}
  {\bibinfo  {journal} {Phys. Rev. ST Accel. Beams}\ }\textbf {\bibinfo
  {volume} {9}},\ \bibinfo {pages} {101301} (\bibinfo {year}
  {2006})}\BibitemShut {NoStop}%
\bibitem [{\citenamefont {Leemans}\ \emph {et~al.}(2014)\citenamefont
  {Leemans}, \citenamefont {Gonsalves}, \citenamefont {Mao}, \citenamefont
  {Nakamura}, \citenamefont {Benedetti}, \citenamefont {Schroeder},
  \citenamefont {T\'oth}, \citenamefont {Daniels}, \citenamefont
  {Mittelberger}, \citenamefont {Bulanov}, \citenamefont {Vay}, \citenamefont
  {Geddes},\ and\ \citenamefont {Esarey}}]{LEEMANS:2014}%
  \BibitemOpen
  \bibfield  {author} {\bibinfo {author} {\bibfnamefont {W.~P.}\ \bibnamefont
  {Leemans}}, \bibinfo {author} {\bibfnamefont {A.~J.}\ \bibnamefont
  {Gonsalves}}, \bibinfo {author} {\bibfnamefont {H.-S.}\ \bibnamefont {Mao}},
  \bibinfo {author} {\bibfnamefont {K.}~\bibnamefont {Nakamura}}, \bibinfo
  {author} {\bibfnamefont {C.}~\bibnamefont {Benedetti}}, \bibinfo {author}
  {\bibfnamefont {C.~B.}\ \bibnamefont {Schroeder}}, \bibinfo {author}
  {\bibfnamefont {C.}~\bibnamefont {T\'oth}}, \bibinfo {author} {\bibfnamefont
  {J.}~\bibnamefont {Daniels}}, \bibinfo {author} {\bibfnamefont {D.~E.}\
  \bibnamefont {Mittelberger}}, \bibinfo {author} {\bibfnamefont {S.~S.}\
  \bibnamefont {Bulanov}}, \bibinfo {author} {\bibfnamefont {J.-L.}\
  \bibnamefont {Vay}}, \bibinfo {author} {\bibfnamefont {C.~G.~R.}\
  \bibnamefont {Geddes}},\ and\ \bibinfo {author} {\bibfnamefont
  {E.}~\bibnamefont {Esarey}},\ }\bibfield  {title} {\bibinfo {title}
  {Multi-gev electron beams from capillary-discharge-guided subpetawatt laser
  pulses in the self-trapping regime},\ }\href
  {https://doi.org/10.1103/PhysRevLett.113.245002} {\bibfield  {journal}
  {\bibinfo  {journal} {Physical Review Letters}\ }\textbf {\bibinfo {volume}
  {113}},\ \bibinfo {pages} {245002} (\bibinfo {year} {2014})}\BibitemShut
  {NoStop}%
\bibitem [{\citenamefont {Chen}\ \emph {et~al.}(2012)\citenamefont {Chen},
  \citenamefont {Esarey}, \citenamefont {Schroeder}, \citenamefont {Geddes},\
  and\ \citenamefont {Leemans}}]{CHEN:2012}%
  \BibitemOpen
  \bibfield  {author} {\bibinfo {author} {\bibfnamefont {M.}~\bibnamefont
  {Chen}}, \bibinfo {author} {\bibfnamefont {E.}~\bibnamefont {Esarey}},
  \bibinfo {author} {\bibfnamefont {C.~B.}\ \bibnamefont {Schroeder}}, \bibinfo
  {author} {\bibfnamefont {C.~G.~R.}\ \bibnamefont {Geddes}},\ and\ \bibinfo
  {author} {\bibfnamefont {W.~P.}\ \bibnamefont {Leemans}},\ }\bibfield
  {title} {\bibinfo {title} {{Theory of ionization-induced trapping in
  laser-plasma accelerators}},\ }\href {https://doi.org/10.1063/1.3689922}
  {\bibfield  {journal} {\bibinfo  {journal} {Physics of Plasmas}\ }\textbf
  {\bibinfo {volume} {19}},\ \bibinfo {pages} {033101} (\bibinfo {year}
  {2012})}\BibitemShut {NoStop}%
\bibitem [{\citenamefont {C.~Li}(2013)}]{LI:2013}%
  \BibitemOpen
  \bibfield  {author} {\bibinfo {author} {\bibfnamefont {C.~T.}\ \bibnamefont
  {C.~Li}, \bibfnamefont {H.~Zha}},\ }\bibfield  {title} {\bibinfo {title}
  {{Calculation of Wakefield in Plasma-Filled Dielectric Capillaries Generated
  by a Relativistic Electron Beam}},\ }in\ \href@noop {} {\emph {\bibinfo
  {booktitle} {Proc. IPAC'13}}},\ \bibinfo {series and number} {International
  Particle Accelerator Conference}\ (\bibinfo  {publisher} {JACoW Publishing,
  Geneva, Switzerland},\ \bibinfo {year} {2013})\BibitemShut {NoStop}%
\bibitem [{\citenamefont {Biagioni}\ \emph {et~al.}(2018)\citenamefont
  {Biagioni}, \citenamefont {Anania}, \citenamefont {Bellaveglia},
  \citenamefont {Brentegani}, \citenamefont {Castorina}, \citenamefont
  {Chiadroni}, \citenamefont {Cianchi}, \citenamefont {{Di Giovenale}},
  \citenamefont {{Di Pirro}}, \citenamefont {Fares}, \citenamefont
  {Ficcadenti}, \citenamefont {Filippi}, \citenamefont {Ferrario},
  \citenamefont {Mostacci}, \citenamefont {Pompili}, \citenamefont {Scifo},
  \citenamefont {Spataro}, \citenamefont {Vaccarezza}, \citenamefont {Villa},\
  and\ \citenamefont {Zigler}}]{BIAGIONI:2018}%
  \BibitemOpen
  \bibfield  {author} {\bibinfo {author} {\bibfnamefont {A.}~\bibnamefont
  {Biagioni}}, \bibinfo {author} {\bibfnamefont {M.}~\bibnamefont {Anania}},
  \bibinfo {author} {\bibfnamefont {M.}~\bibnamefont {Bellaveglia}}, \bibinfo
  {author} {\bibfnamefont {E.}~\bibnamefont {Brentegani}}, \bibinfo {author}
  {\bibfnamefont {G.}~\bibnamefont {Castorina}}, \bibinfo {author}
  {\bibfnamefont {E.}~\bibnamefont {Chiadroni}}, \bibinfo {author}
  {\bibfnamefont {A.}~\bibnamefont {Cianchi}}, \bibinfo {author} {\bibfnamefont
  {D.}~\bibnamefont {{Di Giovenale}}}, \bibinfo {author} {\bibfnamefont
  {G.}~\bibnamefont {{Di Pirro}}}, \bibinfo {author} {\bibfnamefont
  {H.}~\bibnamefont {Fares}}, \bibinfo {author} {\bibfnamefont
  {L.}~\bibnamefont {Ficcadenti}}, \bibinfo {author} {\bibfnamefont
  {F.}~\bibnamefont {Filippi}}, \bibinfo {author} {\bibfnamefont
  {M.}~\bibnamefont {Ferrario}}, \bibinfo {author} {\bibfnamefont
  {A.}~\bibnamefont {Mostacci}}, \bibinfo {author} {\bibfnamefont
  {R.}~\bibnamefont {Pompili}}, \bibinfo {author} {\bibfnamefont
  {J.}~\bibnamefont {Scifo}}, \bibinfo {author} {\bibfnamefont
  {B.}~\bibnamefont {Spataro}}, \bibinfo {author} {\bibfnamefont
  {C.}~\bibnamefont {Vaccarezza}}, \bibinfo {author} {\bibfnamefont
  {F.}~\bibnamefont {Villa}},\ and\ \bibinfo {author} {\bibfnamefont
  {A.}~\bibnamefont {Zigler}},\ }\bibfield  {title} {\bibinfo {title} {Wake
  fields effects in dielectric capillary},\ }\href
  {https://doi.org/https://doi.org/10.1016/j.nima.2018.01.028} {\bibfield
  {journal} {\bibinfo  {journal} {Nuclear Instruments and Methods in Physics
  Research Section A: Accelerators, Spectrometers, Detectors and Associated
  Equipment}\ }\textbf {\bibinfo {volume} {909}},\ \bibinfo {pages} {247}
  (\bibinfo {year} {2018})},\ \bibinfo {note} {3rd European Advanced
  Accelerator Concepts workshop (EAAC2017)}\BibitemShut {NoStop}%
\bibitem [{\citenamefont {Whittum}\ \emph {et~al.}(1991)\citenamefont
  {Whittum}, \citenamefont {Sharp}, \citenamefont {Yu}, \citenamefont {Lampe},\
  and\ \citenamefont {Joyce}}]{WHITTUM:1991}%
  \BibitemOpen
  \bibfield  {author} {\bibinfo {author} {\bibfnamefont {D.~H.}\ \bibnamefont
  {Whittum}}, \bibinfo {author} {\bibfnamefont {W.~M.}\ \bibnamefont {Sharp}},
  \bibinfo {author} {\bibfnamefont {S.~S.}\ \bibnamefont {Yu}}, \bibinfo
  {author} {\bibfnamefont {M.}~\bibnamefont {Lampe}},\ and\ \bibinfo {author}
  {\bibfnamefont {G.}~\bibnamefont {Joyce}},\ }\bibfield  {title} {\bibinfo
  {title} {Electron-hose instability in the ion-focused regime},\ }\href
  {https://doi.org/10.1103/PhysRevLett.67.991} {\bibfield  {journal} {\bibinfo
  {journal} {Physical Review Letters}\ }\textbf {\bibinfo {volume} {67}},\
  \bibinfo {pages} {991} (\bibinfo {year} {1991})}\BibitemShut {NoStop}%
\bibitem [{\citenamefont {Nechaeva}\ \emph {et~al.}(2024)\citenamefont
  {Nechaeva}, \citenamefont {Verra}, \citenamefont {Pucek}, \citenamefont
  {Ranc}, \citenamefont {Bergamaschi}, \citenamefont {Zevi Della~Porta},
  \citenamefont {Muggli}, \citenamefont {Agnello}, \citenamefont {Ahdida},
  \citenamefont {Amoedo}, \citenamefont {Andrebe}, \citenamefont {Apsimon},
  \citenamefont {Apsimon}, \citenamefont {Arnesano}, \citenamefont {Bencini},
  \citenamefont {Blanchard}, \citenamefont {Burrows}, \citenamefont
  {Buttensch\"on}, \citenamefont {Caldwell}, \citenamefont {Chung},
  \citenamefont {Cooke}, \citenamefont {Davut}, \citenamefont {Demeter},
  \citenamefont {Dexter}, \citenamefont {Doebert}, \citenamefont {Farmer},
  \citenamefont {Fasoli}, \citenamefont {Fonseca}, \citenamefont {Furno},
  \citenamefont {Granados}, \citenamefont {Granetzny}, \citenamefont
  {Graubner}, \citenamefont {Grulke}, \citenamefont {Gschwendtner},
  \citenamefont {Guran}, \citenamefont {Henderson}, \citenamefont {Kedves},
  \citenamefont {Kim}, \citenamefont {Kraus}, \citenamefont {Krupa},
  \citenamefont {Lefevre}, \citenamefont {Liang}, \citenamefont {Liu},
  \citenamefont {Lopes}, \citenamefont {Lotov}, \citenamefont
  {Martinez~Calderon}, \citenamefont {Mazzoni}, \citenamefont {Moon},
  \citenamefont {Morales~Guzm\'an}, \citenamefont {Moreira}, \citenamefont
  {Okhotnikov}, \citenamefont {Pakuza}, \citenamefont {Pannell}, \citenamefont
  {Pardons}, \citenamefont {Pepitone}, \citenamefont {Poimenidou},
  \citenamefont {Pukhov}, \citenamefont {Rey}, \citenamefont {Rossel},
  \citenamefont {Saberi}, \citenamefont {Schmitz}, \citenamefont {Senes},
  \citenamefont {Silva}, \citenamefont {Silva}, \citenamefont {Spear},
  \citenamefont {Stollberg}, \citenamefont {Sublet}, \citenamefont {Swain},
  \citenamefont {Topaloudis}, \citenamefont {Torrado}, \citenamefont {Turner},
  \citenamefont {Velotti}, \citenamefont {Verzilov}, \citenamefont {Vieira},
  \citenamefont {Welsch}, \citenamefont {Wendt}, \citenamefont {Wing},
  \citenamefont {Wolfenden}, \citenamefont {Woolley}, \citenamefont {Xia},
  \citenamefont {Yarygova},\ and\ \citenamefont {Zepp}}]{NECHAEVA:2024}%
  \BibitemOpen
  \bibfield  {author} {\bibinfo {author} {\bibfnamefont {T.}~\bibnamefont
  {Nechaeva}}, \bibinfo {author} {\bibfnamefont {L.}~\bibnamefont {Verra}},
  \bibinfo {author} {\bibfnamefont {J.}~\bibnamefont {Pucek}}, \bibinfo
  {author} {\bibfnamefont {L.}~\bibnamefont {Ranc}}, \bibinfo {author}
  {\bibfnamefont {M.}~\bibnamefont {Bergamaschi}}, \bibinfo {author}
  {\bibfnamefont {G.}~\bibnamefont {Zevi Della~Porta}}, \bibinfo {author}
  {\bibfnamefont {P.}~\bibnamefont {Muggli}}, \bibinfo {author} {\bibfnamefont
  {R.}~\bibnamefont {Agnello}}, \bibinfo {author} {\bibfnamefont {C.~C.}\
  \bibnamefont {Ahdida}}, \bibinfo {author} {\bibfnamefont {C.}~\bibnamefont
  {Amoedo}}, \bibinfo {author} {\bibfnamefont {Y.}~\bibnamefont {Andrebe}},
  \bibinfo {author} {\bibfnamefont {O.}~\bibnamefont {Apsimon}}, \bibinfo
  {author} {\bibfnamefont {R.}~\bibnamefont {Apsimon}}, \bibinfo {author}
  {\bibfnamefont {J.~M.}\ \bibnamefont {Arnesano}}, \bibinfo {author}
  {\bibfnamefont {V.}~\bibnamefont {Bencini}}, \bibinfo {author} {\bibfnamefont
  {P.}~\bibnamefont {Blanchard}}, \bibinfo {author} {\bibfnamefont {P.~N.}\
  \bibnamefont {Burrows}}, \bibinfo {author} {\bibfnamefont {B.}~\bibnamefont
  {Buttensch\"on}}, \bibinfo {author} {\bibfnamefont {A.}~\bibnamefont
  {Caldwell}}, \bibinfo {author} {\bibfnamefont {M.}~\bibnamefont {Chung}},
  \bibinfo {author} {\bibfnamefont {D.~A.}\ \bibnamefont {Cooke}}, \bibinfo
  {author} {\bibfnamefont {C.}~\bibnamefont {Davut}}, \bibinfo {author}
  {\bibfnamefont {G.}~\bibnamefont {Demeter}}, \bibinfo {author} {\bibfnamefont
  {A.~C.}\ \bibnamefont {Dexter}}, \bibinfo {author} {\bibfnamefont
  {S.}~\bibnamefont {Doebert}}, \bibinfo {author} {\bibfnamefont
  {J.}~\bibnamefont {Farmer}}, \bibinfo {author} {\bibfnamefont
  {A.}~\bibnamefont {Fasoli}}, \bibinfo {author} {\bibfnamefont
  {R.}~\bibnamefont {Fonseca}}, \bibinfo {author} {\bibfnamefont
  {I.}~\bibnamefont {Furno}}, \bibinfo {author} {\bibfnamefont
  {E.}~\bibnamefont {Granados}}, \bibinfo {author} {\bibfnamefont
  {M.}~\bibnamefont {Granetzny}}, \bibinfo {author} {\bibfnamefont
  {T.}~\bibnamefont {Graubner}}, \bibinfo {author} {\bibfnamefont
  {O.}~\bibnamefont {Grulke}}, \bibinfo {author} {\bibfnamefont
  {E.}~\bibnamefont {Gschwendtner}}, \bibinfo {author} {\bibfnamefont
  {E.}~\bibnamefont {Guran}}, \bibinfo {author} {\bibfnamefont
  {J.}~\bibnamefont {Henderson}}, \bibinfo {author} {\bibfnamefont {M.~A.}\
  \bibnamefont {Kedves}}, \bibinfo {author} {\bibfnamefont {S.-Y.}\
  \bibnamefont {Kim}}, \bibinfo {author} {\bibfnamefont {F.}~\bibnamefont
  {Kraus}}, \bibinfo {author} {\bibfnamefont {M.}~\bibnamefont {Krupa}},
  \bibinfo {author} {\bibfnamefont {T.}~\bibnamefont {Lefevre}}, \bibinfo
  {author} {\bibfnamefont {L.}~\bibnamefont {Liang}}, \bibinfo {author}
  {\bibfnamefont {S.}~\bibnamefont {Liu}}, \bibinfo {author} {\bibfnamefont
  {N.}~\bibnamefont {Lopes}}, \bibinfo {author} {\bibfnamefont
  {K.}~\bibnamefont {Lotov}}, \bibinfo {author} {\bibfnamefont
  {M.}~\bibnamefont {Martinez~Calderon}}, \bibinfo {author} {\bibfnamefont
  {S.}~\bibnamefont {Mazzoni}}, \bibinfo {author} {\bibfnamefont
  {K.}~\bibnamefont {Moon}}, \bibinfo {author} {\bibfnamefont {P.~I.}\
  \bibnamefont {Morales~Guzm\'an}}, \bibinfo {author} {\bibfnamefont
  {M.}~\bibnamefont {Moreira}}, \bibinfo {author} {\bibfnamefont
  {N.}~\bibnamefont {Okhotnikov}}, \bibinfo {author} {\bibfnamefont
  {C.}~\bibnamefont {Pakuza}}, \bibinfo {author} {\bibfnamefont
  {F.}~\bibnamefont {Pannell}}, \bibinfo {author} {\bibfnamefont
  {A.}~\bibnamefont {Pardons}}, \bibinfo {author} {\bibfnamefont
  {K.}~\bibnamefont {Pepitone}}, \bibinfo {author} {\bibfnamefont
  {E.}~\bibnamefont {Poimenidou}}, \bibinfo {author} {\bibfnamefont
  {A.}~\bibnamefont {Pukhov}}, \bibinfo {author} {\bibfnamefont
  {S.}~\bibnamefont {Rey}}, \bibinfo {author} {\bibfnamefont {R.}~\bibnamefont
  {Rossel}}, \bibinfo {author} {\bibfnamefont {H.}~\bibnamefont {Saberi}},
  \bibinfo {author} {\bibfnamefont {O.}~\bibnamefont {Schmitz}}, \bibinfo
  {author} {\bibfnamefont {E.}~\bibnamefont {Senes}}, \bibinfo {author}
  {\bibfnamefont {F.}~\bibnamefont {Silva}}, \bibinfo {author} {\bibfnamefont
  {L.}~\bibnamefont {Silva}}, \bibinfo {author} {\bibfnamefont
  {B.}~\bibnamefont {Spear}}, \bibinfo {author} {\bibfnamefont
  {C.}~\bibnamefont {Stollberg}}, \bibinfo {author} {\bibfnamefont
  {A.}~\bibnamefont {Sublet}}, \bibinfo {author} {\bibfnamefont
  {C.}~\bibnamefont {Swain}}, \bibinfo {author} {\bibfnamefont
  {A.}~\bibnamefont {Topaloudis}}, \bibinfo {author} {\bibfnamefont
  {N.}~\bibnamefont {Torrado}}, \bibinfo {author} {\bibfnamefont
  {M.}~\bibnamefont {Turner}}, \bibinfo {author} {\bibfnamefont
  {F.}~\bibnamefont {Velotti}}, \bibinfo {author} {\bibfnamefont
  {V.}~\bibnamefont {Verzilov}}, \bibinfo {author} {\bibfnamefont
  {J.}~\bibnamefont {Vieira}}, \bibinfo {author} {\bibfnamefont
  {C.}~\bibnamefont {Welsch}}, \bibinfo {author} {\bibfnamefont
  {M.}~\bibnamefont {Wendt}}, \bibinfo {author} {\bibfnamefont
  {M.}~\bibnamefont {Wing}}, \bibinfo {author} {\bibfnamefont {J.}~\bibnamefont
  {Wolfenden}}, \bibinfo {author} {\bibfnamefont {B.}~\bibnamefont {Woolley}},
  \bibinfo {author} {\bibfnamefont {G.}~\bibnamefont {Xia}}, \bibinfo {author}
  {\bibfnamefont {V.}~\bibnamefont {Yarygova}},\ and\ \bibinfo {author}
  {\bibfnamefont {M.}~\bibnamefont {Zepp}} (\bibinfo {collaboration} {AWAKE
  Collaboration}),\ }\bibfield  {title} {\bibinfo {title} {Hosing of a long
  relativistic particle bunch in plasma},\ }\href
  {https://doi.org/10.1103/PhysRevLett.132.075001} {\bibfield  {journal}
  {\bibinfo  {journal} {Phys. Rev. Lett.}\ }\textbf {\bibinfo {volume} {132}},\
  \bibinfo {pages} {075001} (\bibinfo {year} {2024})}\BibitemShut {NoStop}%
\bibitem [{\citenamefont {Ferrario}\ \emph {et~al.}(2013)\citenamefont
  {Ferrario}, \citenamefont {Alesini}, \citenamefont {Anania}, \citenamefont
  {Bacci}, \citenamefont {Bellaveglia}, \citenamefont {Bogdanov}, \citenamefont
  {Boni}, \citenamefont {Castellano}, \citenamefont {Chiadroni}, \citenamefont
  {Cianchi}, \citenamefont {Dabagov}, \citenamefont {{De Martinis}},
  \citenamefont {{Di Giovenale}}, \citenamefont {{Di Pirro}}, \citenamefont
  {Dosselli}, \citenamefont {Drago}, \citenamefont {Esposito}, \citenamefont
  {Faccini}, \citenamefont {Gallo}, \citenamefont {Gambaccini}, \citenamefont
  {Gatti}, \citenamefont {Gatti}, \citenamefont {Ghigo}, \citenamefont
  {Giulietti}, \citenamefont {Ligidov}, \citenamefont {Londrillo},
  \citenamefont {Lupi}, \citenamefont {Mostacci}, \citenamefont {Pace},
  \citenamefont {Palumbo}, \citenamefont {Petrillo}, \citenamefont {Pompili},
  \citenamefont {Rossi}, \citenamefont {Serafini}, \citenamefont {Spataro},
  \citenamefont {Tomassini}, \citenamefont {Turchetti}, \citenamefont
  {Vaccarezza}, \citenamefont {Villa}, \citenamefont {Dattoli}, \citenamefont
  {{Di Palma}}, \citenamefont {Giannessi}, \citenamefont {Petralia},
  \citenamefont {Ronsivalle}, \citenamefont {Spassovsky}, \citenamefont
  {Surrenti}, \citenamefont {Gizzi}, \citenamefont {Labate}, \citenamefont
  {Levato},\ and\ \citenamefont {Rau}}]{FERRARIO:2013}%
  \BibitemOpen
  \bibfield  {author} {\bibinfo {author} {\bibfnamefont {M.}~\bibnamefont
  {Ferrario}}, \bibinfo {author} {\bibfnamefont {D.}~\bibnamefont {Alesini}},
  \bibinfo {author} {\bibfnamefont {M.}~\bibnamefont {Anania}}, \bibinfo
  {author} {\bibfnamefont {A.}~\bibnamefont {Bacci}}, \bibinfo {author}
  {\bibfnamefont {M.}~\bibnamefont {Bellaveglia}}, \bibinfo {author}
  {\bibfnamefont {O.}~\bibnamefont {Bogdanov}}, \bibinfo {author}
  {\bibfnamefont {R.}~\bibnamefont {Boni}}, \bibinfo {author} {\bibfnamefont
  {M.}~\bibnamefont {Castellano}}, \bibinfo {author} {\bibfnamefont
  {E.}~\bibnamefont {Chiadroni}}, \bibinfo {author} {\bibfnamefont
  {A.}~\bibnamefont {Cianchi}}, \bibinfo {author} {\bibfnamefont
  {S.}~\bibnamefont {Dabagov}}, \bibinfo {author} {\bibfnamefont
  {C.}~\bibnamefont {{De Martinis}}}, \bibinfo {author} {\bibfnamefont
  {D.}~\bibnamefont {{Di Giovenale}}}, \bibinfo {author} {\bibfnamefont
  {G.}~\bibnamefont {{Di Pirro}}}, \bibinfo {author} {\bibfnamefont
  {U.}~\bibnamefont {Dosselli}}, \bibinfo {author} {\bibfnamefont
  {A.}~\bibnamefont {Drago}}, \bibinfo {author} {\bibfnamefont
  {A.}~\bibnamefont {Esposito}}, \bibinfo {author} {\bibfnamefont
  {R.}~\bibnamefont {Faccini}}, \bibinfo {author} {\bibfnamefont
  {A.}~\bibnamefont {Gallo}}, \bibinfo {author} {\bibfnamefont
  {M.}~\bibnamefont {Gambaccini}}, \bibinfo {author} {\bibfnamefont
  {C.}~\bibnamefont {Gatti}}, \bibinfo {author} {\bibfnamefont
  {G.}~\bibnamefont {Gatti}}, \bibinfo {author} {\bibfnamefont
  {A.}~\bibnamefont {Ghigo}}, \bibinfo {author} {\bibfnamefont
  {D.}~\bibnamefont {Giulietti}}, \bibinfo {author} {\bibfnamefont
  {A.}~\bibnamefont {Ligidov}}, \bibinfo {author} {\bibfnamefont
  {P.}~\bibnamefont {Londrillo}}, \bibinfo {author} {\bibfnamefont
  {S.}~\bibnamefont {Lupi}}, \bibinfo {author} {\bibfnamefont {A.}~\bibnamefont
  {Mostacci}}, \bibinfo {author} {\bibfnamefont {E.}~\bibnamefont {Pace}},
  \bibinfo {author} {\bibfnamefont {L.}~\bibnamefont {Palumbo}}, \bibinfo
  {author} {\bibfnamefont {V.}~\bibnamefont {Petrillo}}, \bibinfo {author}
  {\bibfnamefont {R.}~\bibnamefont {Pompili}}, \bibinfo {author} {\bibfnamefont
  {A.}~\bibnamefont {Rossi}}, \bibinfo {author} {\bibfnamefont
  {L.}~\bibnamefont {Serafini}}, \bibinfo {author} {\bibfnamefont
  {B.}~\bibnamefont {Spataro}}, \bibinfo {author} {\bibfnamefont
  {P.}~\bibnamefont {Tomassini}}, \bibinfo {author} {\bibfnamefont
  {G.}~\bibnamefont {Turchetti}}, \bibinfo {author} {\bibfnamefont
  {C.}~\bibnamefont {Vaccarezza}}, \bibinfo {author} {\bibfnamefont
  {F.}~\bibnamefont {Villa}}, \bibinfo {author} {\bibfnamefont
  {G.}~\bibnamefont {Dattoli}}, \bibinfo {author} {\bibfnamefont
  {E.}~\bibnamefont {{Di Palma}}}, \bibinfo {author} {\bibfnamefont
  {L.}~\bibnamefont {Giannessi}}, \bibinfo {author} {\bibfnamefont
  {A.}~\bibnamefont {Petralia}}, \bibinfo {author} {\bibfnamefont
  {C.}~\bibnamefont {Ronsivalle}}, \bibinfo {author} {\bibfnamefont
  {I.}~\bibnamefont {Spassovsky}}, \bibinfo {author} {\bibfnamefont
  {V.}~\bibnamefont {Surrenti}}, \bibinfo {author} {\bibfnamefont
  {L.}~\bibnamefont {Gizzi}}, \bibinfo {author} {\bibfnamefont
  {L.}~\bibnamefont {Labate}}, \bibinfo {author} {\bibfnamefont
  {T.}~\bibnamefont {Levato}},\ and\ \bibinfo {author} {\bibfnamefont
  {J.}~\bibnamefont {Rau}},\ }\bibfield  {title} {\bibinfo {title} {{SPARC\_LAB
  present and future}},\ }\href
  {https://doi.org/https://doi.org/10.1016/j.nimb.2013.03.049} {\bibfield
  {journal} {\bibinfo  {journal} {Nuclear Instruments and Methods in Physics
  Research Section B: Beam Interactions with Materials and Atoms}\ }\textbf
  {\bibinfo {volume} {309}},\ \bibinfo {pages} {183} (\bibinfo {year}
  {2013})}\BibitemShut {NoStop}%
\bibitem [{\citenamefont {Alesini}\ \emph {et~al.}(2006)\citenamefont
  {Alesini}, \citenamefont {{Di Pirro}}, \citenamefont {Ficcadenti},
  \citenamefont {Mostacci}, \citenamefont {Palumbo}, \citenamefont
  {Rosenzweig},\ and\ \citenamefont {Vaccarezza}}]{ALESINI:2006}%
  \BibitemOpen
  \bibfield  {author} {\bibinfo {author} {\bibfnamefont {D.}~\bibnamefont
  {Alesini}}, \bibinfo {author} {\bibfnamefont {G.}~\bibnamefont {{Di Pirro}}},
  \bibinfo {author} {\bibfnamefont {L.}~\bibnamefont {Ficcadenti}}, \bibinfo
  {author} {\bibfnamefont {A.}~\bibnamefont {Mostacci}}, \bibinfo {author}
  {\bibfnamefont {L.}~\bibnamefont {Palumbo}}, \bibinfo {author} {\bibfnamefont
  {J.}~\bibnamefont {Rosenzweig}},\ and\ \bibinfo {author} {\bibfnamefont
  {C.}~\bibnamefont {Vaccarezza}},\ }\bibfield  {title} {\bibinfo {title} {Rf
  deflector design and measurements for the longitudinal and transverse phase
  space characterization at sparc},\ }\href
  {https://doi.org/https://doi.org/10.1016/j.nima.2006.07.050} {\bibfield
  {journal} {\bibinfo  {journal} {Nuclear Instruments and Methods in Physics
  Research Section A: Accelerators, Spectrometers, Detectors and Associated
  Equipment}\ }\textbf {\bibinfo {volume} {568}},\ \bibinfo {pages} {488}
  (\bibinfo {year} {2006})}\BibitemShut {NoStop}%
\bibitem [{\citenamefont {Qian}\ \emph {et~al.}(2010)\citenamefont {Qian},
  \citenamefont {Ren}, \citenamefont {Wang}, \citenamefont {Zhang},\ and\
  \citenamefont {Wei}}]{QIAN:2010}%
  \BibitemOpen
  \bibfield  {author} {\bibinfo {author} {\bibfnamefont {M.}~\bibnamefont
  {Qian}}, \bibinfo {author} {\bibfnamefont {C.}~\bibnamefont {Ren}}, \bibinfo
  {author} {\bibfnamefont {D.}~\bibnamefont {Wang}}, \bibinfo {author}
  {\bibfnamefont {J.}~\bibnamefont {Zhang}},\ and\ \bibinfo {author}
  {\bibfnamefont {G.}~\bibnamefont {Wei}},\ }\bibfield  {title} {\bibinfo
  {title} {{Stark broadening measurement of the electron density in an
  atmospheric pressure argon plasma jet with double-power electrodes}},\ }\href
  {https://doi.org/10.1063/1.3330717} {\bibfield  {journal} {\bibinfo
  {journal} {Journal of Applied Physics}\ }\textbf {\bibinfo {volume} {107}},\
  \bibinfo {pages} {063303} (\bibinfo {year} {2010})}\BibitemShut {NoStop}%
\bibitem [{\citenamefont {Shpakov}\ \emph {et~al.}(2019)\citenamefont
  {Shpakov}, \citenamefont {Anania}, \citenamefont {Bellaveglia}, \citenamefont
  {Biagioni}, \citenamefont {Bisesto}, \citenamefont {Cardelli}, \citenamefont
  {Cesarini}, \citenamefont {Chiadroni}, \citenamefont {Cianchi}, \citenamefont
  {Costa}, \citenamefont {Croia}, \citenamefont {Del~Dotto}, \citenamefont
  {Di~Giovenale}, \citenamefont {Diomede}, \citenamefont {Ferrario},
  \citenamefont {Filippi}, \citenamefont {Giribono}, \citenamefont {Lollo},
  \citenamefont {Marongiu}, \citenamefont {Martinelli}, \citenamefont
  {Mostacci}, \citenamefont {Piersanti}, \citenamefont {Di~Pirro},
  \citenamefont {Pompili}, \citenamefont {Romeo}, \citenamefont {Scifo},
  \citenamefont {Vaccarezza}, \citenamefont {Villa},\ and\ \citenamefont
  {Zigler}}]{SHPAKOV:2019}%
  \BibitemOpen
  \bibfield  {author} {\bibinfo {author} {\bibfnamefont {V.}~\bibnamefont
  {Shpakov}}, \bibinfo {author} {\bibfnamefont {M.~P.}\ \bibnamefont {Anania}},
  \bibinfo {author} {\bibfnamefont {M.}~\bibnamefont {Bellaveglia}}, \bibinfo
  {author} {\bibfnamefont {A.}~\bibnamefont {Biagioni}}, \bibinfo {author}
  {\bibfnamefont {F.}~\bibnamefont {Bisesto}}, \bibinfo {author} {\bibfnamefont
  {F.}~\bibnamefont {Cardelli}}, \bibinfo {author} {\bibfnamefont
  {M.}~\bibnamefont {Cesarini}}, \bibinfo {author} {\bibfnamefont
  {E.}~\bibnamefont {Chiadroni}}, \bibinfo {author} {\bibfnamefont
  {A.}~\bibnamefont {Cianchi}}, \bibinfo {author} {\bibfnamefont
  {G.}~\bibnamefont {Costa}}, \bibinfo {author} {\bibfnamefont
  {M.}~\bibnamefont {Croia}}, \bibinfo {author} {\bibfnamefont
  {A.}~\bibnamefont {Del~Dotto}}, \bibinfo {author} {\bibfnamefont
  {D.}~\bibnamefont {Di~Giovenale}}, \bibinfo {author} {\bibfnamefont
  {M.}~\bibnamefont {Diomede}}, \bibinfo {author} {\bibfnamefont
  {M.}~\bibnamefont {Ferrario}}, \bibinfo {author} {\bibfnamefont
  {F.}~\bibnamefont {Filippi}}, \bibinfo {author} {\bibfnamefont
  {A.}~\bibnamefont {Giribono}}, \bibinfo {author} {\bibfnamefont
  {V.}~\bibnamefont {Lollo}}, \bibinfo {author} {\bibfnamefont
  {M.}~\bibnamefont {Marongiu}}, \bibinfo {author} {\bibfnamefont
  {V.}~\bibnamefont {Martinelli}}, \bibinfo {author} {\bibfnamefont
  {A.}~\bibnamefont {Mostacci}}, \bibinfo {author} {\bibfnamefont
  {L.}~\bibnamefont {Piersanti}}, \bibinfo {author} {\bibfnamefont
  {G.}~\bibnamefont {Di~Pirro}}, \bibinfo {author} {\bibfnamefont
  {R.}~\bibnamefont {Pompili}}, \bibinfo {author} {\bibfnamefont
  {S.}~\bibnamefont {Romeo}}, \bibinfo {author} {\bibfnamefont
  {J.}~\bibnamefont {Scifo}}, \bibinfo {author} {\bibfnamefont
  {C.}~\bibnamefont {Vaccarezza}}, \bibinfo {author} {\bibfnamefont
  {F.}~\bibnamefont {Villa}},\ and\ \bibinfo {author} {\bibfnamefont
  {A.}~\bibnamefont {Zigler}},\ }\bibfield  {title} {\bibinfo {title}
  {Longitudinal phase-space manipulation with beam-driven plasma wakefields},\
  }\href {https://doi.org/10.1103/PhysRevLett.122.114801} {\bibfield  {journal}
  {\bibinfo  {journal} {Physical Review Letters}\ }\textbf {\bibinfo {volume}
  {122}},\ \bibinfo {pages} {114801} (\bibinfo {year} {2019})}\BibitemShut
  {NoStop}%
\bibitem [{sup()}]{suppl}%
  \BibitemOpen
  \href@noop {} {\bibinfo  {journal} {See Supplemental Material at [URL will be
  inserted by publisher] for details on the experimental setup}\ }\BibitemShut
  {NoStop}%
\bibitem [{\citenamefont {Chen}(1987)}]{CHEN:1987}%
  \BibitemOpen
\bibfield  {journal} {  }\bibfield  {author} {\bibinfo {author} {\bibfnamefont
  {P.}~\bibnamefont {Chen}},\ }\bibfield  {title} {\bibinfo {title} {A possible
  final focusing mechanism for linear colliders},\ }\href
  {http://inis.iaea.org/search/search.aspx?orig_q=RN:18041971} {\bibfield
  {journal} {\bibinfo  {journal} {Particle Accelerators}\ }\textbf {\bibinfo
  {volume} {20}},\ \bibinfo {pages} {171} (\bibinfo {year} {1987})}\BibitemShut
  {NoStop}%
\bibitem [{\citenamefont {Barov}\ and\ \citenamefont
  {Rosenzweig}(1994)}]{BAROV:1994}%
  \BibitemOpen
  \bibfield  {author} {\bibinfo {author} {\bibfnamefont {N.}~\bibnamefont
  {Barov}}\ and\ \bibinfo {author} {\bibfnamefont {J.~B.}\ \bibnamefont
  {Rosenzweig}},\ }\bibfield  {title} {\bibinfo {title} {Propagation of short
  electron pulses in underdense plasmas},\ }\href
  {https://doi.org/10.1103/PhysRevE.49.4407} {\bibfield  {journal} {\bibinfo
  {journal} {Phys. Rev. E}\ }\textbf {\bibinfo {volume} {49}},\ \bibinfo
  {pages} {4407} (\bibinfo {year} {1994})}\BibitemShut {NoStop}%
\bibitem [{\citenamefont {Verra}\ \emph {et~al.}(2023)\citenamefont {Verra},
  \citenamefont {Gschwendtner},\ and\ \citenamefont {Muggli}}]{VERRA:2022}%
  \BibitemOpen
  \bibfield  {author} {\bibinfo {author} {\bibfnamefont {L.}~\bibnamefont
  {Verra}}, \bibinfo {author} {\bibfnamefont {E.}~\bibnamefont
  {Gschwendtner}},\ and\ \bibinfo {author} {\bibfnamefont {P.}~\bibnamefont
  {Muggli}},\ }\bibfield  {title} {\bibinfo {title} {Focusing of a long
  relativistic proton bunch in underdense plasma},\ }\bibfield  {booktitle}
  {\emph {\bibinfo {booktitle} {2022 IEEE Advanced Accelerator Concepts
  Workshop (AAC)}},\ }\bibfield  {journal} {\bibinfo  {journal} {arXiv preprint
  arXiv:2302.04051}\ }\href {https://doi.org/10.48550/arxiv.2302.04051}
  {10.48550/arxiv.2302.04051} (\bibinfo {year} {2023})\BibitemShut {NoStop}%
\bibitem [{\citenamefont {{R. W. Assmann, R. W. et al.}}(2020)}]{ASSMAN:2020}%
  \BibitemOpen
  \bibfield  {author} {\bibinfo {author} {\bibnamefont {{R. W. Assmann, R. W.
  et al.}}},\ }\bibfield  {title} {\bibinfo {title} {Eupraxia conceptual design
  report},\ }\href {https://doi.org/10.1140/epjst/e2020-000127-8} {\bibfield
  {journal} {\bibinfo  {journal} {The European Physical Journal Special
  Topics}\ }\textbf {\bibinfo {volume} {229}},\ \bibinfo {pages} {3675}
  (\bibinfo {year} {2020})}\BibitemShut {NoStop}%
\end{thebibliography}%
\end{document}